\documentclass[journal]{IEEEtran}
\usepackage{amsmath,amsfonts}
\usepackage{array}
\usepackage{textcomp}
\usepackage{stfloats}
\usepackage{url}
\usepackage{verbatim}
\usepackage{graphicx}
\usepackage{enumitem}
\usepackage{url}
\usepackage{graphics}
\usepackage{tikz}
\usepackage{multirow}
\usepackage{graphicx}
\usepackage{tabularx}
\usepackage{listings}
\usepackage{enumitem}
\usepackage{makecell}
\usepackage{algpseudocode}
\usepackage{siunitx}

\usepackage{mathtools}
\usepackage{multirow}
\usepackage{algpseudocode}
\usepackage{url}

\usepackage[utf8]{inputenc}
\usepackage{enumitem}
\usepackage{url}
\usepackage{tikz}
\usepackage{caption}
\usepackage{subcaption}
\usepackage{multirow}
\usepackage{enumitem}
\usepackage{tabularx}
\usepackage{multirow}
\usepackage{amsthm}
\usepackage{listings}
\usepackage{float}
\usepackage{comment}
\usepackage{hyperref}
\usepackage{amsmath}
\usepackage[ruled,lined,boxed,linesnumbered]{algorithm2e}
\usepackage{amsthm}
\usepackage{balance}

\usepackage{titlesec}
\titlespacing*{\section}
{0pt}{0.5em}{0.3em}
\titlespacing*{\subsection}
{0pt}{0.5em}{0.3em}
\titlespacing*{\subsubsection}
{0pt}{0.5em}{0.3em}
\usepackage[font=small,skip=0pt]{caption}

\newcounter{reviewercount}
\newcounter{commentcount}

\newcommand{\aeditor}%
  {\bigskip\noindent {\bf COMMENTS OF THE ASSOCIATE EDITOR}%
  \setcounter{commentcount}{0}\par 
}

\usepackage[utf8]{inputenc}
\usepackage{enumitem}
\usepackage{url}
\usepackage{multirow}
\usepackage{graphicx}
\usepackage{enumitem}
\usepackage{tabularx}
\usepackage{multirow}
\usepackage{amsthm}
\usepackage{listings}
\usepackage{float}
\usepackage{comment}
\usepackage[ruled,lined,boxed,linesnumbered]{algorithm2e}

\usepackage{soul}

\usepackage{amsthm}
\theoremstyle{plain}

\theoremstyle{definition}

\newtheoremstyle{claim}
  {\topsep}
  {\topsep}
  {}
  {}
  {\itshape}
  {}
  {.5em}
  {\thmname{#1}\thmnumber{ #2}\thmnote{ (#3)}}

\SetKwInput{KwOutput}{Output}
\SetKwInput{KwInput}{Input} 

\graphicspath{ {FIGS/} }

\graphicspath{ {FIGS/} }

\begin{document}

\title{{\em Hier-RTLMP}: A Hierarchical Automatic Macro Placer for Large-scale Complex IP Blocks}
\author{Andrew~B.~Kahng,~\IEEEmembership{Fellow,~IEEE,}
        Ravi Varadarajan,~\IEEEmembership{Student Member,~IEEE}
        and~Zhiang~Wang,~\IEEEmembership{Student Member,~IEEE}}

\maketitle

\begin{abstract}
\textcolor{black}{
In a typical RTL to GDSII flow, floorplanning or macro placement is a critical 
step in achieving decent quality of results {\em (QoR)}. Moreover, in 
today's physical synthesis flows (e.g., Synopsys Fusion Compiler or 
Cadence Genus iSpatial), a floorplan .def with macro and IO pin placements 
is typically needed as an input to the front-end physical synthesis.
Recently, with the increasing complexity of IP blocks, and in particular with
auto-generated RTL for machine learning {\em (ML)} accelerators, the number 
of macros in a single RTL block can easily 
run into the several hundreds.
This makes the task of generating an 
automatic floorplan (.def) with IO pin and macro placements for front-end physical 
synthesis even more critical and challenging. 
The so-called {\em peripheral} approach of forcing macros to the periphery of 
the layout is no longer viable when the ratio of the sum of the macro perimeters 
to the floorplan perimeter is large, since this increases the required 
stacking depth of macros.  
In this paper, we develop a novel multilevel physical planning approach that
exploits the hierarchy and dataflow inherent in the design RTL, and describe
its realization in a new hierarchical macro placer, {\em Hier-RTLMP}. 
{\em Hier-RTLMP} borrows from traditional approaches used in manual 
system-on-chip  (SoC) floorplanning to create an automatic macro placement 
for use with large IP blocks containing very large numbers of macros.
Empirical studies demonstrate substantial improvements
over the previous {\em RTL-MP} macro placement approach \cite{RTL-MP},
and promising post-route improvements relative to a leading commercial 
place-and-route tool. 
}
\end{abstract}

\section{INTRODUCTION}
\label{sec:intro}
\textcolor{black}{
In today's design flows, macro placement is typically performed by expert
human designers.  The resulting floorplan, with placed macros and pin 
locations, is used for both front-end physical synthesis and back-end P\&R flows. 
The human designers use their knowledge of the RTL dataflow and perform 
grouping of the memories based on functionality, along with tiling of the 
macro groups.
\textcolor{black}{Historically, 
due to routing blockages within macros, 
macro tiling is typically done along the periphery of 
the floorplan outline.}
Such a {\em peripheral} methodology works quite well as long as the 
floorplan outline is not too large and the number of macros is limited,
such that macro tiling/stacking depth does not become excessive.
Peripheral placement can also be essential when routing resources over
macros is limited, e.g., when the total number of routing layers is small.
Grouping and tiling of macros further improves power planning and the 
layout of the power grids in the block.}

\textcolor{black}{
However, today's nanometer-era technology nodes have given rise to 
back-end-of-line stacks with more layers,
and auto-generated RTL designs with complex logical hierarchies 
and large numbers of macros. In this modern context, the peripheral 
approach is neither feasible (due to increased stacking depth) nor optimal 
(due to a large penalty in wirelength from not following the dataflow topology). 
In this paper, we describe {\em Hier-RTLMP}, which solves these challenges
by allowing macros to migrate to the core of the floorplan, while preserving 
the use of macro grouping and tiling to ease power grid generation.} 

\textcolor{black}{
Conceptually, we transform the logical hierarchy to a physical
hierarchy and define outlines for the modules in the physical hierarchy (i.e.,
{\em physical clusters}). 
Macros are then automatically tiled along the internal boundaries of given
physical cluster outlines. 
{\em Hier-RTLMP} is scalable to large RTL design blocks because its physical hierarchy 
can have multiple levels, based on size thresholds for physical  
clusters at each level in terms of both area and number of macros. 
Essentially, {\em Hier-RTLMP} mimics and extends the approach of an expert 
back-end designer in creating high-quality, routable floorplans. 
Our contributions are summarized as follows.}

\begin{itemize}[noitemsep,topsep=0pt,leftmargin=*]

\item 
\textcolor{black}{
We propose {\em Hier-RTLMP}, that extends
{\em RTL-MP}~\cite{RTL-MP}
to handle large designs with complex RTL hierarchy and even hundreds of macros.  
{\em Hier-RTLMP} is able to tile macro groups
in the core area of the floorplan.
This enables high-quality outcomes for designs with large numbers of macros,
unlike a pure peripheral macro placement approach that
would see excessive stacking 
depth and destruction of the design dataflow. 
In addition, the tiling of macro groups in {\em Hier-RTLMP} allows for grouping of macros with 
similar functional interactions \textcolor{black} {along with} efficient power grid generation.}

\item
\textcolor{black}{
We develop \textcolor{black}{an} {\em autoclustering} engine that transforms 
the logical hierarchy to a physical hierarchy. 
Unlike {\em RTL-MP}~\cite{RTL-MP} where the physical hierarchy is a single level, 
{\em Hier-RTLMP}'s autoclustering engine creates a multilevel physical hierarchy of 
{\em physical clusters}. 
This enables handling of large RTLs with hundreds of macros, 
and allows for placement of macros within the core area.}

\item 
\textcolor{black}{
We develop a novel shaping engine that determines the allowable shapes 
of a given cluster in the physical hierarchy based on the contents of its 
child clusters and the outline of its parent cluster. 
A unique two-stage, bottom-up / top-down process is used to determine the 
allowable shapes for the clusters before macro placement at each level. 
Our macro placer performs placement and shaping of the clusters in the 
physical hierarchy, level by level.}

\item 
\textcolor{black}{
Extensive empirical studies using Cadence Innovus (v21.1) P\&R, to validate 
floorplan routability and downstream PPA metrics, confirm advantages of {\em Hier-RTLMP}.
{\em Hier-RTLMP} has been tested on both \textcolor{black} {open-source} and industrial large designs,
and compared against a 2021 release of a state-of-the-art commercial macro placer and our previous work
{\em RTL-MP} \cite{RTL-MP}.
{\em Hier-RTLMP} outperforms the commercial macro placer and {\em RTL-MP} for almost all the testcases:
(i) compared to the commercial macro placer, {\em Hier-RTLMP} achieves much better timing metrics (WNS and TNS)
measured post-detailed route;
and  (ii) compared to {\em RTL-MP}, we extend {\em Hier-RTLMP} to handle complex designs with hundreds of macros on which
{\em RTL-MP} fails, and reduce
runtime by $13\times$ relative to 
{\em RTL-MP} for testcases which {\em RTL-MP} can handle.
}

\item
\textcolor{black}{
Our implementation is based on the open-source OpenROAD project infrastructure
with permissive open-sourcing of {\em Hier-RTLMP}.
Experimental runscripts \cite{Junkin22} 
are available in the OpenROAD project \cite{openroad-github}
and the {\em MacroPlacement} project \cite{MacroPlacement-github}.
}

\end{itemize}

\textcolor{black}{
\noindent
The rest of this paper is organized as follows. 
Section~\ref{sec:related_works} reviews related work on macro placement. 
Section~\ref{sec:our_approach} describes the outline of our approach. 
Section~\ref{sec:autoclustering} details implementation of multilevel 
autoclustering, which transforms the RTL logical hierarchy into the 
physical hierarchy.
Section~\ref{sec:shape_function} describes 
\textcolor{black}{the}
generation of allowable shapes for physical clusters, which applies bottom-up analysis along with 
Section~\ref{sec:macro_placement} describes the macro placement engine.  
Section~\ref{sec:experiment} describes experimental results,
and Section~\ref{sec:conclusion} concludes the paper and outlines 
future research directions.
}

\section{RELATED WORK}
\label{sec:related_works}

Broadly, previous works on floorplanning and macro placement can 
be classified into 
packing-based methods, analytical methods and, in the recent past, 
ML and reinforcement learning-based methods.  
Packing-based methods rely on the representation of the physical relationships among
modules in the floorplan along with iterative-perturbative techniques to optimize 
them. Analytical approaches use numerical methods to optimize a floorplan layout directly.  
Both of these approaches optimize a customized cost function that captures
area, congestion and timing.

Most macro placers in the research literature focus on legal placement of 
macros, and optimizing wirelength and/or routability - without considering design 
features such as design hierarchy, macro regularity, dataflow, macro 
guidance, pin access and notch area avoidance.
On the other hand, chip experts do pay attention to these design 
features to produce high-quality macro placements.  To automatically 
generate a competitive, closer to human-quality macro placement, 
some recent works have begun to consider these features. 
\cite{LinLW19}-\cite{LinDYCL21} and~\cite{VidalObiolsCPGM19,VidalObiolsCPGM20,
ChuangNACRV10, ChenYCHL08, HsuCHCC13, HsuCHCLCC14}
utilize design hierarchy to guide macro placement.
\cite{KimL08}-\cite{NookalaCLS05} and~\cite{LinDYCL21,
VidalObiolsCPGM19, VidalObiolsCPGM20}
exploit dataflow and/or timing information to improve the quality 
of macro placement.
\cite{LinDYCC19}-\cite{LinDYCL21},
\cite{VidalObiolsCPGM19, VidalObiolsCPGM20} 
and~\cite{LiuCCK19}-\cite{ChangCC17}
reduce macro displacement to honor the macro guidance given 
by placement prototyping, 
and~\cite{TangW01,YanVC14, ChenYCHL08} consider geometrical constraints 
directly. 
\cite{LinDYCC19}-\cite{LinDYCL21} and~\cite{ChenYCHL08}-\cite{ChangCC17}
can handle macro blockages and/or preplaced macros.
\cite{ChangCC17,HsuCHCC13, HsuCHCLCC14}
try to avoid
notches during macro placement to improve routability.
\cite{LinDLYCP19, ChangCC17}
pay special attention to the effect of regular placement of macros.
However, none of these previous works provides for all of the above features.

\textcolor{black}{While}
\textcolor{black}{
our recent work, {\em RTL-MP}~\cite{RTL-MP}, exploits dataflow inherent 
in  the logical hierarchy and provides many of the above desirable features,
it still has limitations. A major drawback of {\em RTL-MP} is that 
it is not scalable: when there are hundreds of macros, 
its strategy of forcing the macros to the periphery to avoid 
routability issues incurs high QoR penalties.} 

\textcolor{black}{
Last, we note that the Google Brain reinforcement learning-based 
approach to macro placement \cite{MirhoseiniGYJS21} has stimulated 
a great deal of interest across academia and industry. 
As part of a broad discussion of the method and its replication,
macro-heavy open-source benchmark designs in open enablements, along with
implementations of missing or binarized code elements from \cite{CircuitTraining},
have been released via a public GitHub repository~\cite{MacroPlacement-github}. 
\textcolor{black}{
We apply {\em Hier-RTLMP} to the {\em Ariane-133} (NG45, GF12)
testcase from \cite{MacroPlacement-github}
in Section \ref{sec:experiment} below.}
Table~\ref{tab:previousWork} summarizes the main differences between 
our {\em Hier-RTLMP} method and previous works.}

\begin{table}[!htb]
    \caption{\small
    \textcolor{black}{
    Comparison with previous methods. 
    Columns 2-8 respectively indicate use of design hierarchy; 
    use of regular placement of macros; use of dataflow and/or timing; 
    handling of macro guidance (preferred location
    or regions); handling of pin access; handling of 
   macro blockages and/or preplaced macros; and handling of notches. 
   Column 9 indicates the use of multi-level hierarchy to place macros in the 
   core area.}}
    \label{tab:previousWork}
     \resizebox{1.0\columnwidth}{!} {
    \centering
    \begin{tabular}{|c|c|c|c|c|c|c|c|c|}
    \hline
     Methods  
     &  \begin{tabular}{c}
          Design \\
          Hier \\
    \end{tabular}
    & \begin{tabular}{c}
        Shape \\
        Reg \\
    \end{tabular}
    & \begin{tabular}{c}
        Dataflow \\
        Timing \\
    \end{tabular}
    & \begin{tabular}{c}
         Macro \\
         Guide \\
    \end{tabular}
    & \begin{tabular}{c}
         Pin \\
         Access  \\
    \end{tabular}
    & \begin{tabular}{c} 
        Macro \\
        Blockage \\
    \end{tabular}
    & \begin{tabular}{c}
       Notch \\
       Align \\
    \end{tabular} 
    & Multilevel
    \\ \hline
    \begin{tabular}{c}
    ~\cite{AdyaM03}-\cite{ZhanFS06}
    \cite{DreamPlace4.0},~\cite{AutoDMP}
    \end{tabular} & 
     &  &  &  &  &  &  &  \\ \hline
    \begin{tabular}{c}
    ~\cite{LinLW19,ChuangNACRV10} \\
    \end{tabular} &
    \checkmark &  &  &  &  &  &  & \\ \hline
    \begin{tabular}{c}
     ~\cite{KimL08}-\cite{NookalaCLS05} \\
    \end{tabular} & 
     &  & \checkmark &  &  &  & & \\ \hline
    \begin{tabular}{c}
    ~\cite{TangW01, YanVC14}
    \end{tabular} &
     &  &  & \checkmark &  &  & & \\ \hline
    \begin{tabular}{c}
    ~\cite{LiuCCK19}-\cite{ChiouCCC16} \\
    \end{tabular} &
     &  &  &  &  & \checkmark &  & \\ \hline

    ~\cite{LinDYCC19} & \checkmark &  &  &  &  & \checkmark &  & \\ \hline
    
    ~\cite{HsuCHCC13, HsuCHCLCC14} & 
    \checkmark &  &  &  &  &  & \checkmark & \\ \hline
    
    ~\cite{VidalObiolsCPGM19, VidalObiolsCPGM20} 
    & \checkmark &  &  \checkmark & \checkmark &  &  &   & \\ \hline
    ~\cite{ChenYCHL08} & \checkmark &  &  & \checkmark &  & \checkmark &  & \\ \hline
    ~\cite{ChangCC17} &  & \checkmark &  &  &  & \checkmark & \checkmark  & \\ \hline

    ~\cite{LinDLYCP19} & \checkmark & \checkmark &  & \checkmark &  & \checkmark &  & \\ \hline
    ~\cite{LinDYCL21} & \checkmark &  & \checkmark & \checkmark &  & \checkmark &   &  \\ \hline
    {\em RTL-MP} & \checkmark & \checkmark & \checkmark & \checkmark & \checkmark & \checkmark & \checkmark &  \\ \hline
    {\em Hier-RTLMP} & \checkmark & \checkmark & \checkmark & \checkmark & \checkmark & \checkmark & \checkmark & \checkmark \\ \hline
    \end{tabular}
    }
   
\end{table}



\section{OUR APPROACH}
\label{sec:our_approach}

\textcolor{black}{
A distinguishing aspect of our approach is that (i) we extract the 
dataflow inherent  in the RTL description, and (ii) {\em automatically}
map the RTL logical hierarchy to a multilevel physical
hierarchy by performing hierarchical clustering based on the size 
(numbers of standard cells and macros) of each logical module.
The logical-to-physical hierarchy mapping can be one to many: 
a single physical hierarchy cluster can group together multiple logical 
modules from different levels of the logical hierarchy, 
while a single logical module can be split into multiple physical clusters.
Similar to logical hierarchy, the physical hierarchy 
can also be unbalanced. The logical modules at the same level of the 
logical hierarchy may belong to physical clusters at different levels 
of the physical hierarchy.
We use the following terminology, some of which was previously 
introduced in ~\cite{RTL-MP}.}

\begin{figure}
     \centering
     \begin{subfigure}{0.24\textwidth}
         \centering
         \includegraphics[width=\textwidth]{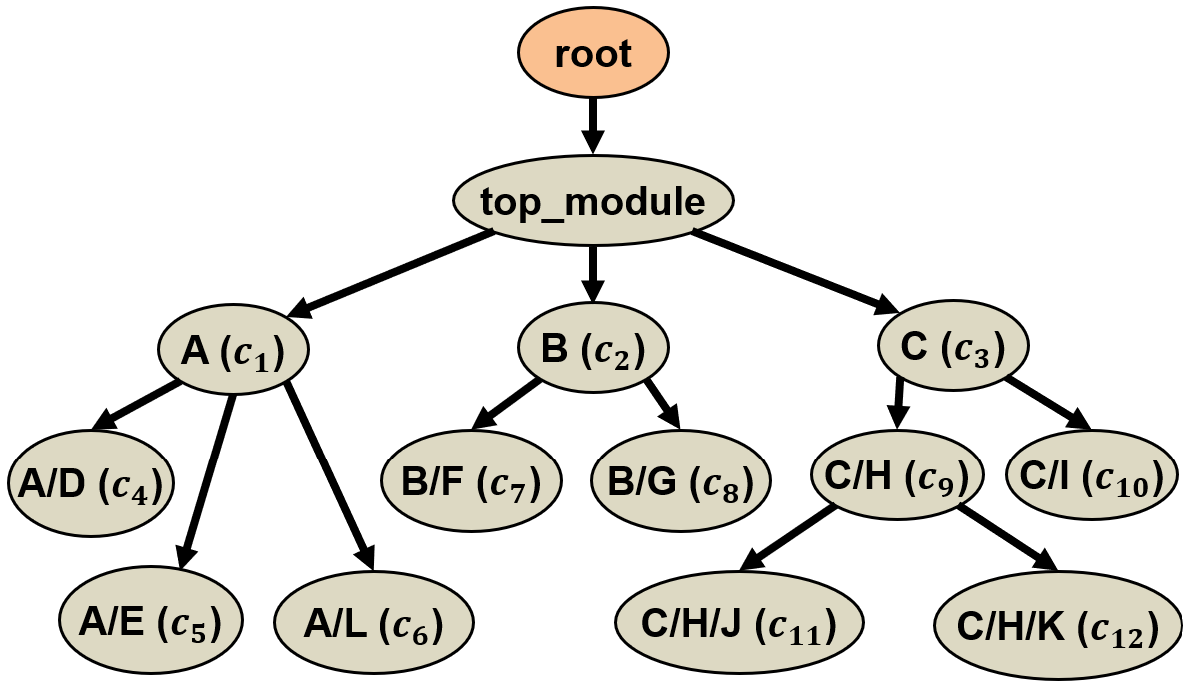}
         \caption{}
     \end{subfigure}
     \hfill
     \begin{subfigure}{0.24\textwidth}
         \centering
         \includegraphics[width=\textwidth]{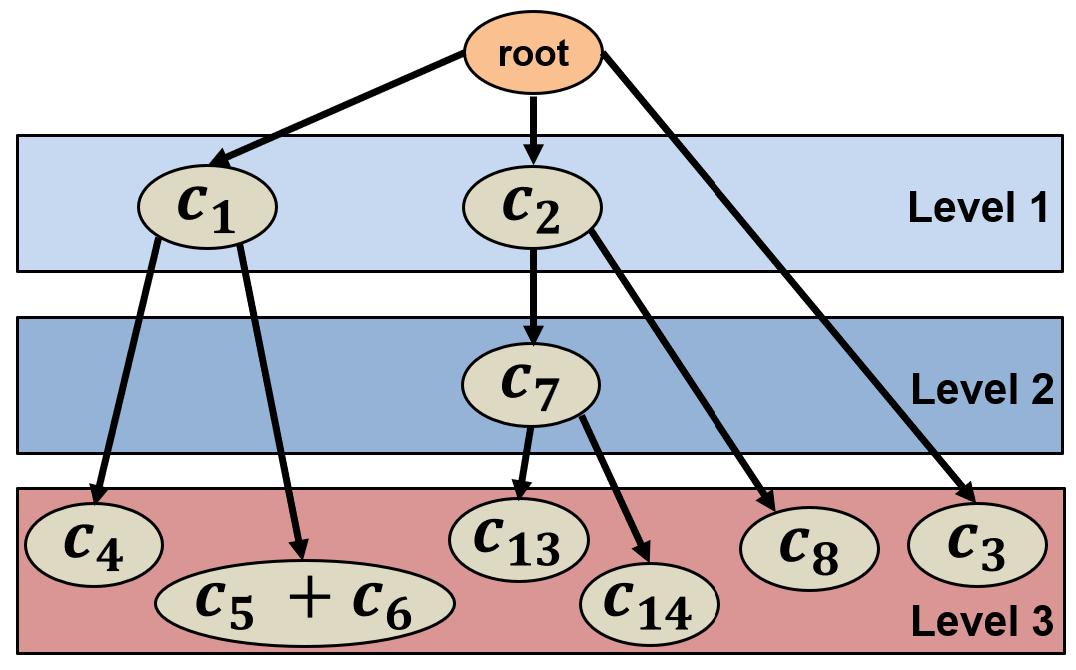}
         \caption{}
     \end{subfigure}
     \hfill
     \begin{subfigure}{0.24\textwidth}
         \centering
         \includegraphics[width=\textwidth]{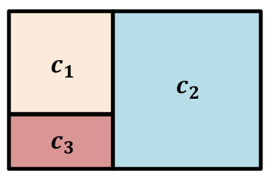}
         \caption{}
     \end{subfigure}
     \begin{subfigure}{0.24\textwidth}
         \centering
         \includegraphics[width=\textwidth]{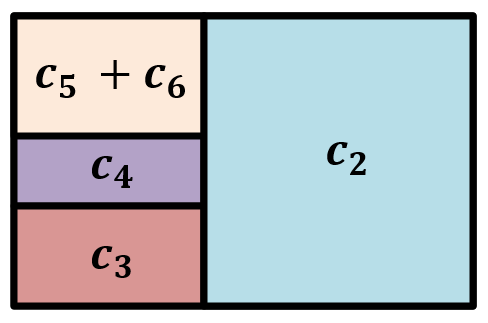}
         \caption{}
     \end{subfigure}
     \hfill
     \begin{subfigure}{0.24\textwidth}
         \centering
         \includegraphics[width=\textwidth]{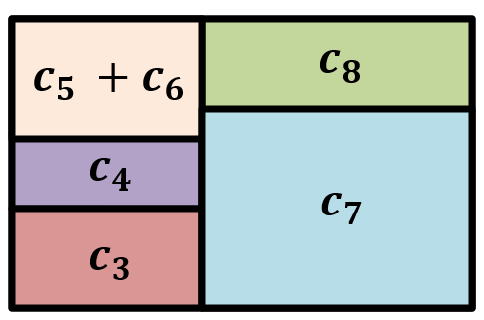}
         \caption{}
     \end{subfigure}
     \begin{subfigure}{0.24\textwidth}
         \centering
         \includegraphics[width=\textwidth]{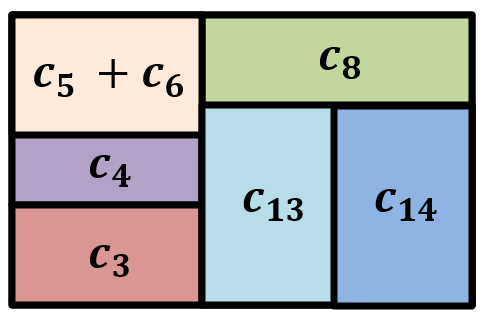}
         \caption{}
     \end{subfigure}
     \caption{
An illustrated example of a hierarchical floorplan scheme. 
(a) Logical hierarchy $T_L$.
(b) Physical hierarchy $T_P$.
In this example, the logical module $C$ is small and contains a limited number of instances. As a result, $c_3$ is classified as a leaf cluster. 
Any cluster within $c_3$ ($c_9$, $c_{10}$, $c_{11}$ and $c_{12}$) is not included in the final $T_P$.
(c) Placed clusters in $T_P$ after visiting $root$ cluster.
(d) Placed clusters in $T_P$ after visiting $c_1$ cluster.
(e) Placed clusters in $T_P$ after visiting $c_2$ cluster. 
(f) Placed clusters in $T_P$ after visiting $c_7$ cluster,
where $c_7$ is dissolved into $c_{13}$ and $c_{14}$.
The placement of clusters is done in a pre-order DFS 
(depth-first search) manner.}
    \label{fig:hier}  
\end{figure}

\begin{itemize}[noitemsep,topsep=0pt,leftmargin=*]
\item \textcolor{black}{
      The {\em logical hierarchy} ($T_L$) is the original RTL hierarchy. 
      Figure \ref{fig:hier}(a) shows an example of logical hierarchy $T_L$.
      Each node in $T_L$ represents a logical module in the netlist.}

\item \textcolor{black}{
      The {\em physical hierarchy} ($T_P$) is the result of autoclustering 
      and defines the physical clusters.}  The physical hierarchy 
      can have multiple levels and is not necessarily balanced. 
      Figure \ref{fig:hier}(b) shows an example of the physical hierarchy $T_P$.

\item \textcolor{black}{
      A {\em physical cluster} ($c$) is a module in the physical hierarchy. 
      It can contain other physical clusters. 
      In Figure \ref{fig:hier}(b), $c_1, c_2, \textcolor{black}{c_3}, c_4, c_5 + c_6, c_7, c_8, c_{13}, c_{14}$ 
      are the physical clusters.  The top level of the design is the root physical cluster.}

\item \textcolor{black}{A {\em macro} is a hard IP block, such as an SRAM block,
which is predesigned and has a fixed layout.
When placing a macro, we can change its location and orientation
(typically, by mirroring and \SI{180}{\degree} rotation, since the orientation
of poly gates cannot be changed)
but we cannot change anything inside it.}

\item A {\em standard-cell cluster} is a physical cluster that contains only standard cells. 

\item A {\em macro cluster} is a physical cluster that contains only macros. 

\item A {\em mixed cluster} is a physical cluster that contains
      both macros and standard cells.       

\item \textcolor{black}{
      A {\em leaf cluster} is a standard-cell cluster or a macro cluster 
      that does not have any children in the physical hierarchy $T_P$.
      In Figure \ref{fig:hier}(b), $c_3$, $c_4$, $c_5 + c_6$, $c_8$, $c_{13}$ and $c_{14}$ are the leaf clusters.}

\item \textcolor{black}{
    A {\em bundled pin} is the physical \textcolor{black}{abstraction} of a group of pins within the same boundary.
    All the pins belonging to a given bundled pin 
    have the same physical location as the bundled pin.
}

\end{itemize}

\begin{figure}[!htb]
    \centering
    \includegraphics[width=0.85\columnwidth]{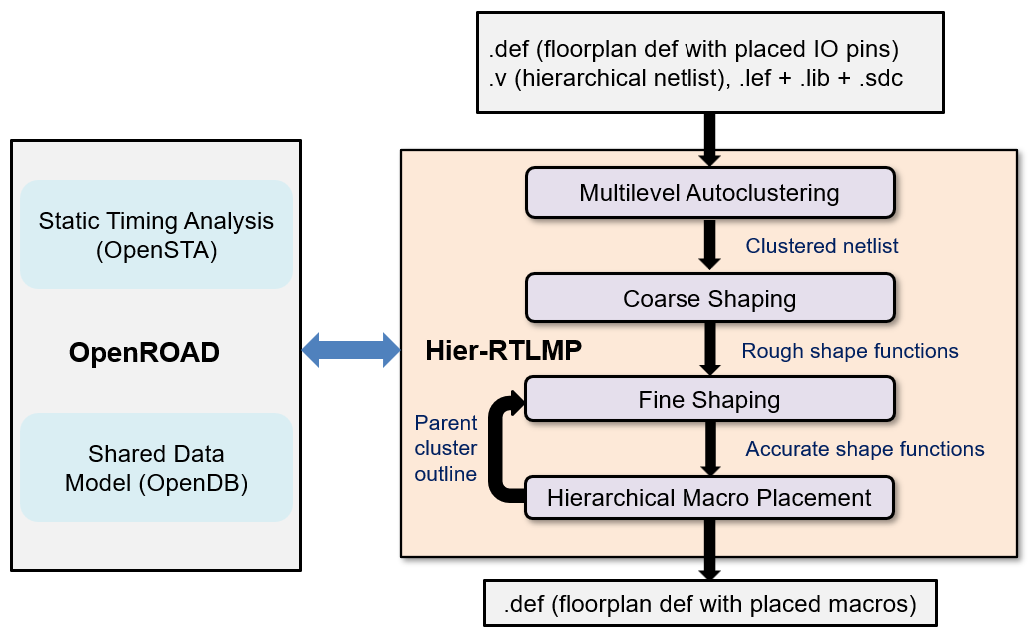}
    \caption{\textcolor{black}{{\em Hier-RTLMP} flow.}}
    \label{fig:Hier-RTLMP}
\end{figure}

\noindent
\textcolor{black}{
{\em Hier-RTLMP} is built on top of the open-source OpenROAD infrastructure~\cite{KahngS21, openroad-github} and works with the
components within the OpenROAD flow, as shown in \textcolor{black}{Figure \ref{fig:Hier-RTLMP}}.
Our previous work, {\em RTL-MP}~\cite{RTL-MP}, partitions the top-level design 
into a set of leaf clusters, resulting in a single level of physical hierarchy.
A known weakness of {\em RTL-MP} is that the single level of hierarchy
makes it very difficult to generate decent macro placements for large,
complex designs with hundreds of macros.
By contrast, {\em Hier-RTLMP's} multilevel physical hierarchy 
framework effectively scales the complexity of design 
blocks that can be handled, and incorporates tiling of macro groups 
along the periphery of physical clusters so as to honor the dataflow and
produce a top-level floorplan with macros in the core area as needed. }

\textcolor{black}{
The fundamental idea of {\em Hier-RTLMP} is that we first create a multilevel
physical hierarchy, then shape and place the clusters in the physical hierarchy,
one level at a time in a \textcolor{black}{pre-order} depth-first search manner.
An overview of our {\em Hier-RTLMP} algorithm is presented in Algorithm~\ref{alg:Hier-RTLMP_new}. 
\textcolor{black}{
The input is a synthesized hierarchical gate-level netlist and 
a floorplan .def file that contains the block outline (width $canvas.width$ and height $canvas.height$), fixed IO pin or pad locations, 
along with macros that have been placed beforehand, if applicable.} 
The output is a floorplan .def file that has legal 
macro placements for \textcolor{black}{macros} 
and region constraints for standard-cell clusters.
{\em Hier-RTLMP} first converts the logical hierarchy $T_L$ of the netlist 
into a  physical hierarchy $T_P$ \textcolor{black}{(Section~\ref{sec:autoclustering})}.
Then, {\em Hier-RTLMP} determines the allowable shape functions for each of the clusters 
in a two-phase bottom-up and top-down manner \textcolor{black}{(Section~\ref{sec:shape_function})}.
Finally, {\em Hier-RTLMP} works top-down from the root physical cluster in a pre-order depth-first search manner, 
and shapes and places the child clusters \textcolor{black}{(Section~\ref{sec:macro_placement})}.}

\begin{algorithm}[!h]
    \small
    \SetKwData{}{left}\SetKwData{This}{this}\SetKwData{Up}{Up}
    \SetKwInOut{Input}{input}\SetKwInOut{Output}{output}
    \KwInput {\textcolor{black}{Hierarchical netlist $N$, Placed IO pins (or PAD) \\
              \quad \quad \quad block outline ($canvas.width$, $canvas.height$)} \\}
    \KwOutput{\textcolor{black}{High-quality floorplan $F_N$ with fixed macro placements\\}}
    \BlankLine{}
    \textcolor{black}{
    initialize the floorplan $F_N$ with $canvas.width$ and $canvas.height$ \\
    update the positions of placed IO pins in $F_N$ \\
    convert the logical hierarchy $T_L$ of $N$ to the physical hierarchy $T_P$ \textcolor{black}{(Section~\ref{sec:autoclustering})} \\
    calculate the rough shape functions for each cluster in a bottom-up manner \textcolor{black}{ (Section~\ref{sec:shape_function})}  \\
    $c_p \gets$ root cluster in $T_P$ \\
    $clusters \gets c_p.GetChildren()$ \\
    adjust the possible shape functions for $clusters$ based on the outline of $c_p$ \textcolor{black}{ (Section~\ref{sec:shape_function})}  \\
    place $clusters$ within the outline of cluster $c_p$ \textcolor{black}{ (Section~\ref{sec:macro_placement})} \\
    update the positions of $clusters$ in $F_N$ \\
    \For{leaf cluster $c_f \in clusters$} {
       place the macros in leaf macro cluster $c_f$ (Section~\ref{sec:macro_placement}) \\
    }
    \For {cluster $c \in c_p.GetChildren()$} {
        $c_p \gets c$; repeat  Lines 5-15
    }
    \Return  $F_N$ \\   
    \caption{{\em Hier-RTLMP}}
    \label{alg:Hier-RTLMP_new}}
\end{algorithm}

\section{MULTILEVEL AUTOCLUSTERING}
\label{sec:autoclustering}

\textcolor{black}{
In this section, we describe the multilevel autoclustering approach.
Subsection \ref{subsec:multilevel-autoclustering}
gives an overview, and 
Subsection \ref{subsec:single-level-autoclustering} provides
a detailed explanation of how we perform autoclustering at each level.}

\subsection{Overview and Algorithm}
\label{subsec:multilevel-autoclustering}

\textcolor{black}{
In today's design flows, 
expert human designers utilize clustering
as an essential pre-processing step for macro placement.}
In this step the structural netlist representation of the design is converted to a clustered netlist in which the nodes are clusters and nets are bundled connections between the clusters.  
The clustering step is typically done by users interactively and in a 
top-down manner, by analyzing logical hierarchy, dataflow, 
connections between macros, IO pins, and 
critical timing paths~\cite{floorplan}. 
Such analysis helps the user to understand the structure
of the design and the dataflow, which provides insights into the 
{\em ideal} locations of the various clusters and macros.
While it is useful for users to perform clustering manually 
and understand the physical implications of the design,
it is also important to have an autoclustering engine
that can fully and automatically generate meaningful clusters.
This is especially true for designs produced by automatic RTL 
generators 
for ML applications; these can have complex RTL structures with 
long, inscrutable auto-generated module names.
We therefore perform autoclustering based on the logical hierarchy
of the design, connection signatures of clusters, and timing hops or 
indirect connections between macros and IO pins as outlined in {\em RTL-MP}~\cite{RTL-MP} -- and extend this to create a multilevel 
physical hierarchy.

The {\em multilevel autoclustering} engine in {\em Hier-RTLMP} extends the autoclustering engine in {\em RTL-MP}~\cite{RTL-MP} to support multiple levels of physical hierarchy.
\textcolor{black}{
It is essential for handling very large RTL blocks with multiple hundreds of macros.}
Moreover, the intermediate levels of physical hierarchy help preserve the global dataflow and allow for the placement of \textcolor{black}{macros} within the core of the floorplan.
The multilevel autoclustering engine first converts
logical hierarchy $T_L$ to physical hierarchy $T_P$ through a one-to-one mapping, i.e., transforming each logical module into a physical cluster directly. 
It then decides which clusters to merge and which clusters to dissolve \textcolor{black}{level by level.}
Conceptually, peer clusters are merged if their individual sizes are below the minimum size threshold for the current level and if they have similar connection patterns to other clusters. 
A cluster is  dissolved \textcolor{black}{into} its child clusters if its size is greater than the maximum size threshold for the current level.

\textcolor{black}{
In contrast to fully following the logical hierarchy, the operations of merging small peer clusters and dissolving large clusters enable us to extract the dataflow and functionality of the RTL. 
For example, some designs may have all the macros in a single logical module, even if these macros have \textcolor{black}{completely} different functionalities and connect to different standard-cell logical modules.
\textcolor{black}{Other designs could have the contents of a bus fabric, 
which connect to different sections of the block,}
in the same module of the logical hierarchy. 
In both these cases, it is obvious that fully following the logical hierarchy to create physical clusters will lead to suboptimal results from both \textcolor{black}{the} timing and congestion \textcolor{black}{perspectives}. 
However, the operations of merging and dissolving enable the multilevel autoclustering engine in {\em Hier-RTLMP} to effectively handle such scenarios from a physical floorplanning perspective.}

\begin{algorithm}[!h]
    \small
    \SetKwData{}{left}\SetKwData{This}{this}\SetKwData{Up}{Up}
    \SetKwInOut{Input}{input}\SetKwInOut{Output}{output}
    \KwInput{\textcolor{black}{Hierarchical netlist $N$, Placed IO pins,} \textcolor{black}{ Maximum depth of the physical hierarchy tree $num\_level$}}
    \KwOutput{\textcolor{black}{Physical hierarchy $T_P$ (depth $\leq  num\_level$)}}
    \textcolor{black}{
    construct the logical hierarchy $T_L$ by traversing $N$ \\
    convert $T_L$ to the physical hierarchy $T_P$ \textcolor{black}{by} mapping each logical module in $T_L$ to a physical cluster $T_P$ \\
    divide each \textcolor{black}{block} boundary into $num\_segment$ segments and create a bundled pin for each segment to represent
    the preplaced IO pins (pads) lying on that segment \\
    model each bundled pin as a child cluster of root cluster $root$ of $T_P$ \\
    $c_p \gets root$ ; $level\_id \gets 1$ \\
    \While{$level\_id <= num\_level$ \textbf{and} $level\_id > 0$} {
       single-level autoclustering($c_p$, $level\_id$) \textcolor{black}{(Section~\ref{subsec:single-level-autoclustering})} \\
       \For {each child cluster $c$ of $c_p$} {
          $c_p \gets c$ ; $level\_id \gets level\_id + 1$ \\
          repeat the while loop [Lines 6-13] \\
       }
       $level\_id \gets level\_id - 1$ \\
    }}
    for each leaf cluster that is a mixed cluster, break it into a standard-cell cluster and a macro cluster \\
    for each leaf cluster that is a macro cluster, mark its macros as single-macro macro clusters and group them based on connection signature. 
    Then, for each newly formed macro cluster with macros of different sizes, 
    break it down and group its macros based on their sizes \\
    update connections between clusters by adding virtual connections between macro clusters and corresponding standard-cell clusters \\
    update connections between clusters by adding virtual connections between clusters based on information flow and number of (latch) hops \\
    \textcolor{black}{
    \Return physical hierarchy $T_P$
    \caption{Multilevel Autoclustering}
    \label{alg:autocluster}}
\end{algorithm}

\textcolor{black}{The algorithm for multilevel autoclustering is shown in 
Algorithm~\ref{alg:autocluster}.}
\textcolor{black}{
The entire autoclustering algorithm can be divided into the following steps.}
    
\noindent    
\textcolor{black}{\textbf{Step 1:} [Lines 1-2]
    We create the logical hierarchy $T_L$ by traversing the hierarchical netlist in a
    \textcolor{black}{pre-order} depth-first search manner. 
    We then transform the logical hierarchy $T_L$ to the physical hierarchy $T_P$ based on a
    one-to-one mapping of each logical module to a physical cluster, i.e., $T_P$ = $T_L$.
    This one-to-one mapping between logical modules and physical clusters
    allows us to follow the logical hierarchy during the process of dissolving and merging physical clusters at each level. 
    In the example of Figure \ref{fig:hier}(a),
    each logical module is mapped to a physical cluster directly, 
    such as B/G to $c_8$ and C/H/J to $c_{11}$.}

\noindent
\textbf{Step 2:} [Lines 3-4]
    We create bundled pins by 
    dividing each boundary edge evenly into $num\_segment$ segments, 
    and assigning each bundled pin to the center of its corresponding segment 
    (Section \ref{sec:experiment} discusses the effect of $num\_segment$).
    Each bundled pin is treated as a physical cluster without physical area,
    and is added as a child cluster of the root node in the physical hierarchy $T_P$.
    A design may have thousands of IO pins.  
    The bundled pin model can reduce the connection complexity significantly, 
    while preserving the connections between clusters and IO pins and the affinity of clusters to related 
    segments of boundaries in the floorplan.

\noindent
\textcolor{black}{\textbf{Step 3:} [Lines 5-13] 
    We transform the physical hierarchy $T_P$ in a pre-order depth-first search manner.
    \textcolor{black}{
    At each level, we create current-level clusters by breaking down the clusters from the parent level
    based on the size thresholds (the number of standard cells and the number of macros in a cluster) of the current level (see Section \ref{subsec:single-level-autoclustering} for details).}
    After this step, we have a physical hierarchy $T_P$ with a depth less than or equal to $num\_level$.
    In the example of Figure \ref{fig:hier}, after Steps 1-3, we convert the 
    logical hierarchy $T_L$ in Figure \ref{fig:hier}(a) into 
    the physical hierarchy $T_P$ ($num\_level$ = 3) in Figure \ref{fig:hier}(b).}
    
\noindent
\textbf{Step 4:} [Line 14]
    \textcolor{black}{Each leaf cluster that is a mixed cluster,
    is broken into a standard-cell cluster and a macro cluster.}
    Separating into standard-cell clusters and macro clusters makes it easier to calculate the shape function for all the clusters (Section~\ref{sec:shape_function}).
    We also add a virtual weighted connection between each macro cluster and its corresponding 
    standard-cell cluster to bias the macro placer to place them together.
    
\noindent
\textbf{Step 5:} [Line 15] For each macro cluster, 
    we mark each of its macros as a {\em single-macro macro cluster} 
    (that is, a macro cluster consisting of only one macro) 
    and group these single-macro macro clusters based on connection signatures 
    (Section~\ref{subsec:single-level-autoclustering}). 
    \textcolor{black}{This re-grouping of macros ensures that macros 
    that have (near-)identical physical connectivity to other clusters are grouped 
    together irrespective of 
    where their instantiations occur in the logical hierarchy.}
    \textcolor{black}{
    If a newly-formed macro cluster has macros with different footprints,
    we break it down and regroup macros within it based on the footprints 
    of the macros, which ensures that all the macros in the same macro cluster have the same shape.}
    Grouping macros based on identical footprints enables
    better tiling of macros in macro groups with
    less wasted whitespace, thereby 
    achieving regular placement of macros.

\begin{figure}
     \centering
         \includegraphics[width=0.45\textwidth]{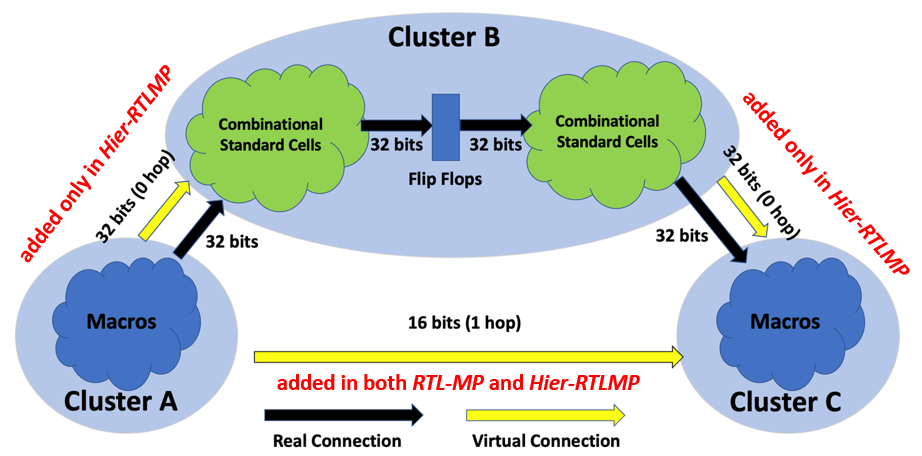}
     \caption{
     \textcolor{black}{
     Virtual connections between clusters. Black arrows 
    represent real connections and yellow arrows represent virtual 
    connections. The virtual connections represented by yellow arrows can help capture critical timing paths between Clusters A, B and C. In {\em RTL-MP}, 
    \textcolor{black}{there is a virtual connection only}
    between macro cluster A and macro cluster C. In {\em Hier-RTLMP}, the virtual connections between all the clusters are captured.}}
   \label{fig:virtualConnection}
\end{figure}

\noindent
\textbf{Step 6:} [Line 17] 
    We add virtual connections between clusters to capture the timing 
    criticality between the clusters and to handle 
    multiple pipeline stages in timing connections between clusters.
    The physical distances between
    components on critical timing paths should be minimized to improve
    performance.  One approach is to determine all the
    critical timing paths (e.g., having negative slack) and overlay
    them on clusters.  This is time-consuming and unnecessary
    since delay and slack calculation are not accurate at the floorplan stage.
    In this work,  we 
    use an approach similar to that of~\cite{VidalObiolsCPGM19, VidalObiolsCPGM20} and 
    define virtual ({\em indirect}) connections as 
    \begin{equation}
    \label{eq:num_hops}
    \resizebox{0.425\textwidth}{!}{
     $virtual\_connections(A, B) = \frac{information\_flow(A, B)} {2^{num\_hops}}$
    }
    \end{equation}
    where {\em information\_flow} corresponds to connection bitwidth 
    and {\em num\_hops} is the length of the shortest path of registers between clusters.
    However, unlike the {\em Dataflow\_Affinity} defined in \cite{VidalObiolsCPGM19,VidalObiolsCPGM20}, 
    we use a more aggressive decaying factor as in~\cite{MirhoseiniGYJS20,MirhoseiniGYJS21} to 
    capture the most important {\em indirect} connections. 
    \textcolor{black}{
    If the register distance ($num\_hops)$ between clusters is greater than $num\_hop\_thr$ (default = 4), 
    then no virtual connection is added.} 
    Also, in most designs, IO pins are registered before connecting to macros.
    It is important to capture this affinity during macro placement to ensure that 
    the macros are placed close to ``connected'' IO pins. 
    By treating each bundled pin as a cluster without physical area, 
    indirect connections from primary inputs 
    and to primary outputs can also be taken into account.
    \textcolor{black}{
    As shown in Figure \ref{fig:virtualConnection},
    in contrast to {\em RTL-MP}~\cite{RTL-MP} which ignores the indirect connections between standard-cell clusters, 
    {\em Hier-RTLMP} considers the indirect connections for all types of clusters (standard-cell clusters, macro clusters, mixed clusters and IOs), thus capturing the timing-critical paths \textcolor{black}{more effectively}.
    \textcolor{black}{To more easily stop the calculation of indirect connections when 
    the $num\_hops$ is larger than $num\_hop\_thr$,
    we start with all the macros and IO pins, 
    and traverse the sequential graph in a breath-first search manner.
    Here we cannot use the topological order because the sequential
    graph of the netlist may contain cycles.}    
    Here, \textcolor{black}{vertices of} the sequential graph consist of all the 
    flip-flops, macros and IO pins, 
    and each edge in the sequential graph represents a 
    \textcolor{black}{directed}
    combinational path,
    \textcolor{black}{i.e., sequential adjacency}
    in the netlist.}

\subsection{Single-level Autoclustering}
\label{subsec:single-level-autoclustering}

\textcolor{black}{
At each level of the autoclustering process, 
we first create current-level clusters by breaking down the clusters 
from the parent level based on the size
threshold of the current level, then merge small peer clusters 
and dissolve large clusters through traversing the logical hierarchy.
We define four parameters related to the size threshold as described in~\cite{RTL-MP}:  
{\em max\_num\_inst} and {\em min\_num\_inst}, 
the maximum and minimum number of standard-cell instances in a cluster; 
\textcolor{black}{and} {\em max\_num\_macro} and {\em min\_num\_macro}, 
the maximum and minimum number of macros in a cluster.
The {\em Hier-RTLMP} flow takes $num\_level$ (the maximum depth of physical hierarchy tree) 
and $level\_ratio$ (the ratio of size threshold between the current level and its parent level) 
as inputs.\footnote{
\textcolor{black}{
$num\_level$ is 2 by default, and $level\_ratio$ is 10 by default. 
On the one hand, the number of clusters at each level should be small enough such that {\em Hier-RTLMP} can effectively place clusters level by level. 
On the other hand, the size of leaf clusters should
\textcolor{black}{not be} \textcolor{black}{too} large, 
such that the ``bundled pin'' for each cluster is still accurate enough for capturing connectivity.
\textcolor{black}{
With this in mind, based on studies of multiple 
designs in different technology nodes,
the default values of $num\_level$ and $level\_ratio$
are respectively set to 2 and 10.}
We also always set $min\_num\_inst$ and $min\_num\_macro$ respectively to  
$max\_num\_inst / 2$ and \textcolor{black} {$max\_num\_macro / 2$.}
}}
Then the size threshold of each level 
\textcolor{black}{decays}
exponentially  as the $level\_id$ increases.
\textcolor{black}{For} example, $max\_num\_inst$ of level $level\_id$ is the total number of instances of the design divided by
$level\_ratio ^ {level\_id}$.}

After determining the size threshold for the current level,
we can create current-level clusters by breaking down the parent clusters from the upper level.  
Since we have created our physical hierarchy
$T_P$ through a one-to-one mapping of each logical module to 
a physical cluster (Algorithm \ref{alg:autocluster}), 
we can further merge small peer clusters and dissolve large clusters 
through traversing the logical hierarchy.

\textcolor{black}{
In the example of Figure~\ref{fig:hier},
logical modules $A$, $B$, $C$, $A/D$, $A/E$, $A/L$, $B/F$ and $B/G$ in logical 
hierarchy $T_L$ respectively correspond to 
physical clusters $c_1$, $c_2$, $c_3$, $c_4$, $c_5$, $c_6$, $c_7$ and $c_8$ in physical hierarchy $T_P$.}
\textcolor{black}{
Here, cluster $c_7$ which corresponds to logical module $B/G$ is 
between the minimum, maximum size thresholds of level $2$,
but larger than the maximum size threshold of level $3$.
Hence $c_7$ is dissolved into clusters $c_{13}$ and $c_{14}$ using a partitioner. 
Clusters $c_5$ and $c_6$ which correspond to logical modules $A/E$ and $A/L$ 
are smaller than the minimum size threshold of level $3$. 
Hence $c_5$ and $c_{6}$ are merged into cluster $c_5 + c_6$.
Cluster $c_3$ which corresponds to logical module
$C$ is within the size thresholds of level $3$.
Hence, $c_3$ is a leaf cluster,
and any cluster within $c_3$ ($c_9$, $c_{10}$, $c_{11}$ and $c_{12}$) is not included in the final $T_P$.}

\begin{algorithm}[!h]
    \small
    \SetKwData{}{left}\SetKwData{This}{this}\SetKwData{Up}{Up}
    \SetKwInOut{Input}{input}\SetKwInOut{Output}{output}
    \KwInput {\textcolor{black}{Parent cluster $c_p$, Level id $level\_id$ \\}}
    \BlankLine{}
    \textcolor{black}{
    determine size thresholds for current level \\
    $children \gets c_p.GetChildren()$ \\
    initialize an empty cluster list $new\_children$ \\
    \uIf {$children.size() == 0$}{
      $new\_children \gets$ recursively call TritonPart to bipartition $c_p$ until
      the number of standard cells in each child cluster is less than $max\_num\_inst$
    }
    \Else {
       initialize an empty cluster list $candidate\_clusters$ \\
       \For{each cluster $c \in children$} {
          \uIf{$c.num\_macro > max\_num\_macro$ \textbf{or}
               $c.num\_std\_cell > max\_num\_inst$
           } {
              $c_p \gets c$; repeat Lines 2-18
           } \uElseIf{$c.num\_macro < min\_num\_macro$ \textbf{and}
                      $c.num\_std\_cell < min\_num\_inst$} {
             $candidate\_clusters.push\_back(c)$ \\
           } \Else{
              $new\_children.push\_back(c)$ \\
           }
        }
        merge clusters in $candidate\_clusters$ based on connection signature \\
    }
    $c_p.SetChildren(new\_children)$ \\    
    \BlankLine{}
    \caption{Single-level Autoclustering}
    \label{alg:autocluster1}}
\end{algorithm}

\noindent
\textcolor{black}{
The detailed algorithm is presented 
in Algorithm \ref{alg:autocluster1} and 
the entire single-level autoclustering algorithm can be divided into the following steps.}

\noindent
\textcolor{black}{\textbf{Step 1:} [Line 1] Determine the size thresholds for the current level, including
    $max\_num\_inst$, $min\_num\_inst$, $max\_num\_macro$ and $min\_num\_macro$.}

\noindent
\textcolor{black}{\textbf{Step 2:} [Lines 4-5] If the parent cluster $c_p$ is a leaf 
    cluster,}\footnote{\textcolor{black}{Actually in this case,  $c_p$ is flat, i.e., this cluster 
    consists of standard cells and macros directly instead of other logical modules.}}
    \textcolor{black}{ 
    and its size is larger than the threshold for the current level,  
    we recursively call the open-source partitioner {\em TritonPart} \cite{openroad-github}} \textcolor{black}{in min-cut mode
    to bipartition the cluster $c_p$
    into child clusters that meet the size threshold (number of standard cells) 
    for the current level.
    Each instance in cluster $c_p$ is treated as a
    vertex.}
    \textcolor{black}{
    Here, we do not need to consider the size threshold for number of 
    macros because the macros will be grouped separately 
    later \textcolor{black}{[Algorithm~\ref{alg:autocluster} Line 15]}.}

\noindent    
\textcolor{black}{\textbf{Step 3:} [Lines 6-18] We handle each child cluster of parent cluster $c_p$ based 
   on the size thresholds 
   in a \textcolor{black}{pre-order} depth-first search manner. During this step, we break down each large cluster
   (number of macros $ > max\_num\_macro$ or number of standard cells $ > max\_num\_inst$) 
   according to \textcolor{black} {the} logical hierarchy\footnote{\textcolor{black}{Since we have created our 
   physical hierarchy $T_P$ through one-to-one mapping of each logical 
   module to a physical cluster (Algorithm \ref{alg:autocluster}), 
   the logical hierarchy has been encoded into the physical hierarchy $T_P$.}},
   and merge the small clusters (number of macros $  < min\_num\_macro$ 
   and number of standard cells $ < min\_num\_inst$) based on {\em connection signature}. 
   \textcolor{black}{Here, the {\em connection signature}} is the connection topology of a cluster with respect 
   to the other clusters in the physical hierarchy $T_P$, 
   which calibrates the connection similarity of clusters.
   More specifically, for a cluster $c$ and a threshold $\epsilon_{net}$,  
   the connection signature $conn\_hash(c)$ of cluster $c$ with respect
   to clusters $\{c_1, c_2, ..., c_n\}$ (termed as {\em reference clusters}) 
   is a vector of size $n$ such that
   \begin{equation*}
    conn\_hash(c)[i] =  \begin{cases}
                       1  \quad  & \text{if the number of nets between} \\
                        & \text{$c$ and $c_i$ is larger than $\epsilon_{net}$}, \\
                       0  \quad &\text{otherwise} \\
                      \end{cases}
    \end{equation*}
    The \textcolor{black}{threshold} $\epsilon_{net}$ (default $= 50$)\footnote{
    \textcolor{black}{
    Our empirical studies have shown that  $\epsilon_{net}$ should not be set based on the Rent parameter of the design and the current level. The purpose of $\epsilon_{net}$ is to ensure the connection patterns are contributed by signal bits instead of common global nets.
    Our empirical studies for, e.g., Ariane (NG45) (Table \ref{tab:hier_rtlmp_result}) show that,
    setting $\epsilon_{net} = 40$ worsens TNS from -55ns to -125ns, while TNS 
    remains at -55ns with $\epsilon_{net} = 60$.}}
    is introduced to remove
    spurious effects of
    common global nets such as 
    scan or reset signals.\footnote{\textcolor{black}{All multiple-pin nets are decomposed using a directed star model.}}
    The intuition behind merging clusters with the same connection signature 
    is that clusters with similar connection patterns would want to stay together in the floorplan.  
    Moreover, the small clusters being merged satisfy the following two criteria:  
    (i) they belong to the same logical module;
    \textcolor{black}{and} (ii) they have similar bus structures, 
    i.e., the number of nets connecting 
    \textcolor{black}{to}
    any reference cluster should be similar.}

\begin{figure}
     \centering
     \begin{subfigure}[b]{0.23\textwidth}
         \centering
         \includegraphics[width=\textwidth]{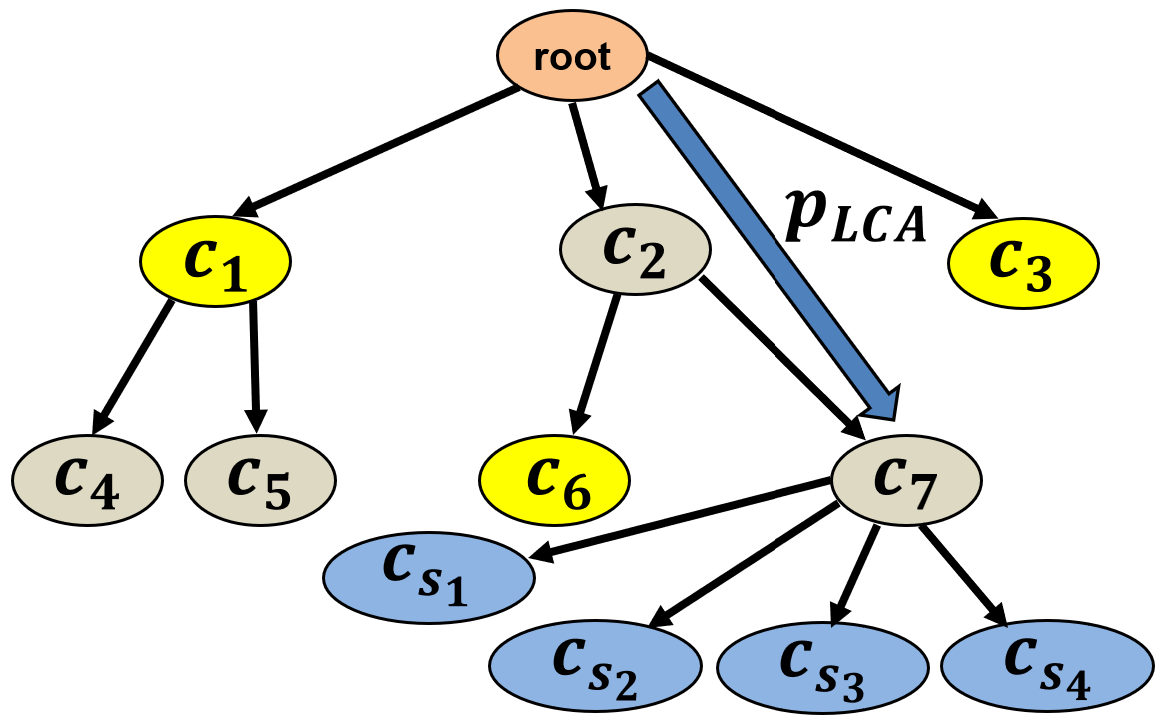}
         \caption{}
     \end{subfigure}
     \hfill
     \begin{subfigure}[b]{0.23\textwidth}
         \centering
         \includegraphics[width=\textwidth]{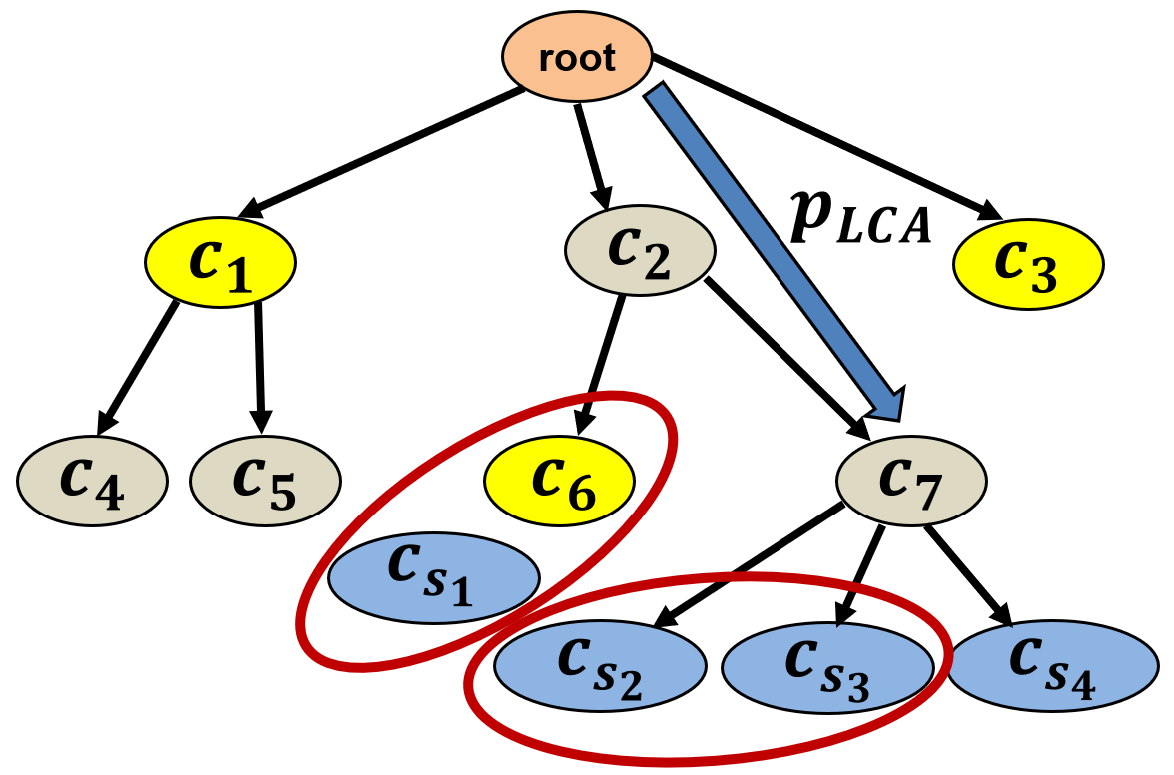}
         \caption{}
     \end{subfigure}
     \hfill
     \begin{subfigure}[b]{0.3\textwidth}
         \centering
         \includegraphics[width=\textwidth]{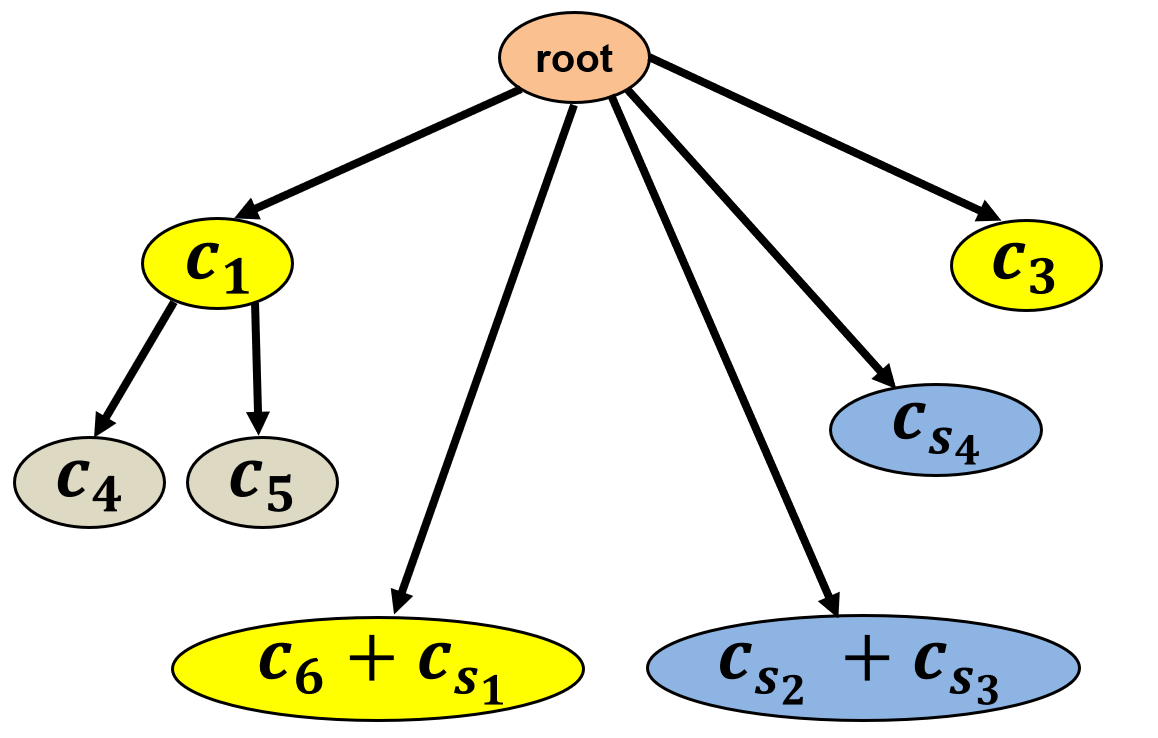}
         \caption{}
     \end{subfigure}
    \caption{
    \textcolor{black}{
    An example of merging clusters based on connection signature.
    $\{ c_{s_1}, c_{s_2}, c_{s_3}, c_{s_4} \}$ are the candidate clusters to be merged.
    $c_7$ is the least common ancestor and $p_{LCA} = \{root, c_2, c_7\}$ the shortest path between root node $T_P$ and $c_7$. 
    $\{ c_1, c_3, c_6 \}$ are the reference clusters.
    Notably, cluster $c_{s1}$ only connects to cluster $c_6$,
    and cluster $c_{s2}$ and cluster $c_{s3}$ have the same connection signature.
    Thus, $c_{s1}$ and $c_6$ will be merged into one cluster $c_{s1} + c_6$,
    and $c_{s2}$ and $c_{s3}$ will be merged into one cluster $c_{s2} + c_{s3}$, as indicated by red ellipses in (b).
    Shown in the figure: physical hierarchy $T_P$ (a) before merging;
    (b) during merging; and
    (c) after merging.
    }} 
    \label{fig:merge}
\end{figure}

\begin{algorithm}[!h]  
    \small
    \SetKwData{}{left}\SetKwData{This}{this}\SetKwData{Up}{Up}
    \SetKwInOut{Input}{input}\SetKwInOut{Output}{output}
    \KwInput {\textcolor{black}{Candidate clusters  $\{ c_{s_1}, ...,  c_{s_k} \}$}}
    \BlankLine{}
    \textcolor{black}{
    find the least common ancestor $v_{LCA}$ of  $\{ c_{s_1}, ...,  c_{s_k} \}$ \textcolor{black}{in} tree $T_P$ \\
    find the shortest path $p_{LCA}$ between root node of $T_P$ and $v_{LCA}$ \\
    $reference\_clusters = \{ \}$ \\
    \For{each node $v \in p_{LCA}$} {
       \If{$v \textcolor{black}{ \neq } v_{LCA}$} {
            \For{each node $u \in v.children()$} {
                \If{$u \not\in p_{LCA}$}{
                    $reference\_clusters.append(u)$ \\
                }   
            }
       }
    }
    calculate the connection signatures of $\{ c_{s_1}, ...,  c_{s_k} \}$ with respect to $reference\_clusters$ \\
    \For {each cluster $c \in  \{ c_{s_1}, ...,  c_{s_k} \}$ }{
       \If{$c$ connects to only one of the reference clusters}{
          merge $c$ with that reference cluster
       }
    }
    merge remaining clusters in $\{ c_{s_1}, ...,  c_{s_k} \}$ with the same connection signature \\
    update tree $T_P$ \\
    \caption{Merge Clusters Based on Connection Signature}
    \label{alg:merge}}
    \end{algorithm}

\noindent
\textcolor{black}{
In contrast to {\em RTL-MP}~\cite{RTL-MP}, 
we extend the calculation of connection signature to a multilevel context.
Details are given in Algorithm~\ref{alg:merge}.
The entire process is illustrated with the example 
shown in Figure~\ref{fig:merge}.
In this example, $\{c_{s_1}, c_{s_2}, c_{s_3}, c_{s_4} \}$ are the candidate clusters 
to be merged. 
First, we find the least common ancestor of $\{ c_{s_1}, c_{s_2}, c_{s_3}, 
c_{s_4} \}$, i.e., $c_7$  [Line 1].
Second,  we determine the shortest path $P_{LCA} = {root, c_2, c_7}$ 
between root node and $c_7$ [Line 2].
Third, we traverse $P_{LCA}$ and \textcolor{black}{identify}
all the reference clusters $\{ c_1, c_3, c_6 \}$ [Lines 3-12].
\textcolor{black}{We then} calculate the connection signatures of 
$\{ c_{s_1}, c_{s_2}, c_{s_3}, c_{s_4}\}$ with respect to $\{ c_1, c_3, c_6 \}$ [Line 13].
Fourth, if a candidate cluster 
\textcolor{black}{connects to only}
one of the reference clusters,
we merge the candidate cluster with that reference cluster directly [Lines 14-18].
\textcolor{black}{As shown in the figure,}
if cluster $c_{s1}$ only connects to cluster $c_6$, we will merge  $c_{s1}$ with $c_6$. 
Finally, we merge remaining candidate clusters \textcolor{black}{having} the same connection signature 
and update the physical hierarchy tree $T_P$ [Lines 19-20].
For example, if cluster $c_{s2}$ and $c_{s3}$ have the same connection signature,
we will merge $c_{s2}$ and $c_{s3}$.  
The final physical hierarchy is shown in Figure~\ref{fig:merge}(c). 
As shown in the figure, the final physical hierarchy is 
different from the original logical hierarchy:
(i) there are physical clusters in the physical hierarchy that correspond 
to multiple logical modules in the logic hierarchy;
\textcolor{black}{and} (ii) the physical hierarchy tree $T_p$ is unbalanced and 
has a different structure from the logical hierarchy.}

\section{\textcolor{black}{SHAPE FUNCTIONS FOR CLUSTERS}}
\label{sec:shape_function}

\textcolor{black}{
Shape function calculation~\cite{KahngLMH-text} determines the allowable 
rectangular shapes for each cluster. 
In {\em RTL-MP}~\cite{RTL-MP}, the physical hierarchy has a single level,
and all clusters are leaf clusters; shape functions are continuous 
for standard-cell clusters and discrete for macro clusters. 
By contrast, in {\em Hier-RTLMP}\textcolor{black}{,} shape functions are calculated for all 
the intermediate levels of the physical hierarchy, leading up to the root, 
using a two-step process. 
First, shape functions are initially calculated bottom-up from leaf clusters. 
Second, once the shape \textcolor{black}{functions} for \textcolor{black}{all  clusters} in the physical hierarchy have been calculated, 
\textcolor{black} {they are}  further refined before the top-down macro placement of the clusters, 
starting from the root cluster. 
Conceptually, during bottom-up determination of the shape functions, 
we still do not know the outline nor the IO pin locations for the parent cluster. 
During the top-down macro placement of the physical clusters, 
starting from the root cluster (i.e., the top-level design), 
both the outline and the IO pin locations are fixed. 
Hence, the shape functions of  child clusters can be further 
refined to better accommodate the parent cluster's outline and IO pin locations. 
This enables improved convergence and outcomes (runtime and QoR) 
during the macro placement of a given parent cluster. 
The remainder of this section gives details of how cluster shape 
functions are determined.}

\begin{figure}[!htb]
    \centering
    \includegraphics[width=0.75\columnwidth]{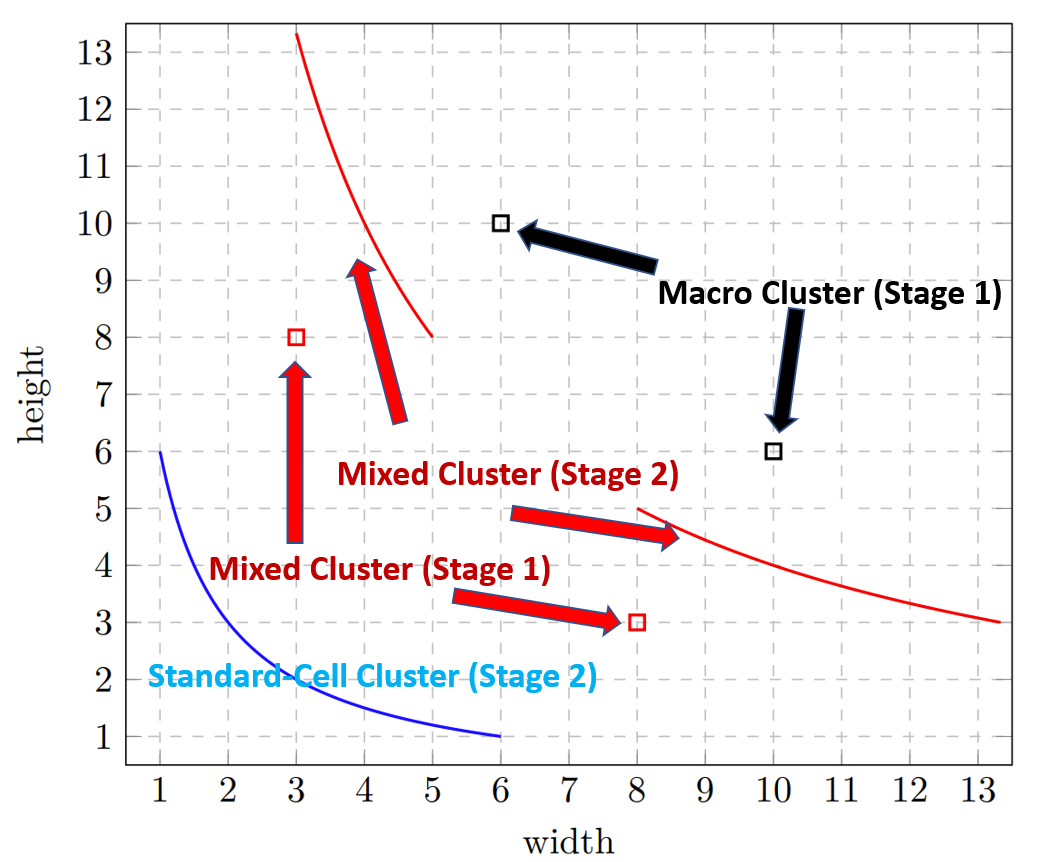}
    \caption{\textcolor{black}{Example shape curves for standard-cell, macro, and
    mixed cluster types,}
    \textcolor{black}{during coarse (stage 1) and fine (stage 2) shaping.
    The blue line represents the shape curve of a standard-cell cluster during fine (stage 2) shaping;
    the two red dots and two segments of red lines respectively
    represent the shape curves of a mixed cluster during coarse
    (stage 1) and fine (stage 2) shaping;
    and the two black dots represent the shape curve of 
    a macro cluster during coarse (stage 1) shaping.}}
    \label{fig:shape}
\end{figure}

\textcolor{black}{
Recall from Section \ref{sec:our_approach} 
above that there are {\em standard-cell}, {\em macro},
and {\em mixed} types of clusters.} 
\begin{itemize}[noitemsep,topsep=0pt,leftmargin=*]
   \item \textcolor{black}{A standard-cell cluster (containing only standard cells) has fixed  area and continuous aspect ratio within  given lower and upper bounds.
   As shown in Figure \ref{fig:shape}, the shape function of a standard-cell cluster is a continuous curve.}
   \item \textcolor{black}{A macro cluster (containing only macros) can have different areas and discrete aspect ratios.
   As shown in Figure \ref{fig:shape}, the shape function of a macro cluster is a set of
   discrete points.}
   \item \textcolor{black}{A mixed cluster (containing both macros and standard cells) has fixed area and 
   piecewise-continuous aspect ratios.
   As shown in Figure \ref{fig:shape}, 
   the shape function of a mixed cluster \textcolor{black}{(red trace)} 
   is a piecewise-continuous curve.}
\end{itemize}

\textcolor{black}{
We calculate the shape functions of clusters in 
a ``$\Lambda$-shaped'' multilevel manner, i.e., following the 
traditional multilevel paradigm.
This enables {\em Hier-RTLMP} to dynamically adjust the possible shapes for
each cluster, based on the placement 
(Section \ref{sec:macro_placement}).\footnote{\textcolor{black}{In our implementation, \textcolor{black}{each shape of a cluster is rectangular
and represented by a pair of values (width, height).}}}
The overall shape function calculation can be divided into two stages, which
respectively perform {\em coarse shaping} and {\em fine shaping}, as follows.}

\noindent
{\bf Coarse Shaping (Stage 1).} 
\textcolor{black}{
During coarse shaping,
we determine the rough shape function for each cluster in a bottom-up manner,
which means the shapes of a cluster are based on the tilings of all its child clusters.
This method of bottom-up calculation for shape functions guarantees that the determined shape of a macro or mixed cluster is always large enough to accommodate the tilings of all the included macros.}
Coarse shaping consists of \textcolor{black}{the} following steps.
\textcolor{black}{First,} the shape of a standard-cell cluster is constrained 
by a user-specified parameter $min\_ar$ (default value = 0.33), 
i.e., the aspect ratio of a standard-cell cluster is in the range $[min\_ar, \frac{1.0}{min\_ar}]$.
\textcolor{black}{Second,} for the shape function of a macro cluster, we use Simulated Annealing to calculate possible macro tilings,
which have minimum area and can fit into the floorplan.
\textcolor{black}{Third,} for the shape function of a mixed cluster, we ignore the area of standard cells,
and use Simulated Annealing to calculate possible tilings of its child clusters.
\textcolor{black}{As applied in this phase, the Sequence Pair-based annealing supports four solution perturbation (move) operators with respective probabilities 0.3, 0.3, 0.3 and 0.1:
\begin{itemize}[noitemsep,topsep=0pt,leftmargin=*]
    \item \textbf{Op1:} Swap two clusters in the first sequence; 
    \item \textbf{Op2:} Swap two clusters in the second sequence; 
    \item \textbf{Op3:} Swap two clusters in each of both sequences; and
    \item \textbf{Op4:} Change the shape of a cluster by randomly picking one shape from its available shapes.
\end{itemize}
The details of Simulated Annealing are as follows:
the number of moves per iteration is 500;
the total number of iteration is 200;
the initial acceptance probability is 0.9;
and the minimum temperature is $1e^{10}$.}
Here we also force the aspect ratio of a mixed cluster \textcolor{black}{to be} within the range $[min\_ar, \frac{1.0}{min\_ar}]$
if there are multiple available possible tilings.
As shown in Figure ~\ref{fig:shape}, after coarse shaping we know the
shape functions for macro clusters, and the discrete approximation of 
the shape functions for mixed clusters, for the entire physical hierarchy.

\noindent
\textcolor{black}{
{\bf Fine Shaping (Stage 2).} Fine shaping is done in a top-down manner, 
starting from the root cluster\textcolor{black}{,} 
and before the macro placement of each given cluster to dynamically adjust the possible shapes of each of the child 
clusters. At this stage, we refine the possible shapes of each cluster based on 
the fixed outline and location of its parent cluster. 
For the root cluster, which is the top-level block, the fixed outline 
and the IO pin locations are the input constraints, usually derived from 
the starting floorplan .def file. Details are given in 
Algorithm \ref{alg:shaping_function_2}. 
First, we remove the area occupied by placement blockages
from the available area of parent cluster \textcolor{black}{$c_p$} [Lines 1-2].
Second, for each macro cluster, we abandon the shapes that cannot fit 
into the outline of parent cluster \textcolor{black}{$c_p$} [Lines 4-7].
Third, for each mixed cluster, we inflate the area of 
standard cells based on the target utilization $util$, then 
convert the discrete shapes into continuous shapes (see the example 
in Figure \ref{fig:shape}) by adding the area of standard 
cells  [Lines 8-21].
Fourth, we check if there is enough empty space for standard-cell 
clusters [Lines 26-28].
Finally, we inflate the area of standard cells based on target dead 
space $t\_dead\_space$.
Here we ignore the area of {\em tiny clusters} which only have tens of standard cells
[Lines 29-38]. The target dead space controls 
the amount of whitespace in the floorplan.}

\textcolor{black}{
Conceptually, the larger \textcolor{black}{the} $t\_dead\_space$ and $util$ 
(\textcolor{black}{the smaller the area of mixed clusters and standard-cell clusters}),
the easier it is to generate a 
valid macro placement that optimizes wirelength, but this may cause significant routing issues.
In order to avoid high congestion, physical clusters with higher macro \textcolor{black}{density} usually
require lower utilization. Thus, in contrast to all previous works, 
we dynamically inflate the area of standard cells in mixed clusters 
and standard-cell clusters based on target utilization $util$ and target 
dead space $t\_dead\_space$, respectively.
Furthermore, since it is difficult to find {\em universal} values 
of target utilization $util$ and target dead space $t\_dead\_space$ parameters
that succeed for all designs, we sweep these two parameters within 
given user-specified computing resource and runtime constraints, and pick 
the best floorplan obtained from the given budget of trials. More 
specifically, for each target utilization $util$, we sweep the target dead 
space $t\_dead\_space$ parameter. The sweeping stops when the current shape 
functions for clusters result in a valid placement of clusters; in our 
experiments, this occurs within 10 to 30 trials.
The runtime reported in Section \ref{sec:experiment} includes the total 
runtime of all the trials.
The runtime analysis in Section \ref{sec:experiment} also suggests that the
runtime overhead of  sweeping the target dead space $t\_dead\_space$
and target utilization $util$ is not significant.}

\begin{algorithm}[!h]
\small
    \SetKwData{}{left}\SetKwData{This}{this}\SetKwData{Up}{Up}
    \SetKwInOut{Input}{input}\SetKwInOut{Output}{output}
    \KwInput {\textcolor{black}{Physical hierarchy tree $T_P$, Placed parent cluster $c_p$, \\
              \quad \quad \quad Target utilization $util$, \\
              \quad \quad \quad Target dead space $t\_dead\_space$, \\
              \quad \quad \quad 
              Minimum aspect ratio $min\_ar$}}
    \BlankLine{}
    \textcolor{black}{
    $avail\_area \gets c_p.GetOutlineArea()$ \\
    $avail\_area \mathrel{{-}{=}}$ area occupied by placement blockages \\
    \For {cluster $c$ $\in$ $c_p.GetChildren()$}{
       \If {$c$ is a macro cluster} {
          abandon the shapes which cannot fit into $c_p$ \\
          $avail\_area \mathrel{{-}{=}} c.GetArea()$ \\
       }
       \If {$c$ is a mixed cluster} {
          abandon the shapes which cannot fit into $c_p$ \\
          $shapes \gets c.GetShapes()$ \\
          $area \gets $ the minimum area of possible shapes \\
          $area \gets area + c.GetStdCellArea() / util$ \\
          initialize an empty list $aspect\_ratios$ \\
          \For {each $shape$ in $shapes$} {
          $ar = [\frac{shape.height}{area / shape.height},
             \frac{area / shape.width}{shape.width}]$ \\
             $aspect\_ratios.push\_back(ar)$ \\
          }
          $c.SetArea(area)$ \\
          \textcolor{black}{$c.SetAspectRatios(aspect\_ratios)$} \\
          $avail\_area \mathrel{{-}{=}} area $ \\
       }
    }
    \If {$c_p.GetStdCellArea() == 0.0$} {
       \Return true \\
    }
    \If {$avail\_area \leq 0.0$} {
       \Return false \\
    }
    $inflat\_ratio = \frac{c_p.GetStdCellArea()}{avail\_area * (1 - t\_dead\_space)} $ \\
    \For{each child standard-cell cluster $c$ of $c_p$} {
        \If{$c$ is a tiny cluster}{
          $c.SetArea(0.0)$ \\
        } 
        \Else {
          $c.SetArea(\frac{c.GetArea()}{inflat\_ratio})$ \\
          $c.SetAspectRatios([min\_ar, \frac{1.0}{min\_ar}])$ \\
        }    
    }
\Return true \\
\caption{Fine Shaping}
\label{alg:shaping_function_2}}
\end{algorithm}

\section{HIERARCHICAL MACRO PLACEMENT}
\label{sec:macro_placement}

\textcolor{black}{
In this section, we describe our approach to top-down hierarchical macro placement. 
Subsection \ref{sec:macroPlacementEngine} describes how we place and shape the physical clusters, 
one level at a time, in a  pre-order depth-first search manner.
Subsection \ref{sec:pinAlignmentEngine} describes how we determine the 
location and orientation of macros in each leaf macro cluster.}

\subsection{Placement of Clusters}
\label{sec:macroPlacementEngine}

\textcolor{black}{
We shape and place the physical clusters, one level at a time, in a pre-order 
depth-first search manner starting from the root cluster.
Before the placement of clusters at each level,
we first determine the shape functions for all clusters 
based on the outline and location of the parent cluster 
(Fine \textcolor{black}{Shaping} in Section~\ref{sec:shape_function}).
For example, in Figure \ref{fig:hier},
before we place \textcolor{black} {and shape} clusters $c_7$ and $c_8$,
we adjust the shape functions of $c_7$ and $c_8$ based on the outline and location
of their parent cluster $c_2$.
We then calculate the connections between clusters, IO pins and other clusters of the parent level.
\textcolor{black}{The} other clusters of the parent level are \textcolor{black}{the} {\em reference clusters} described in Section \ref{subsec:single-level-autoclustering}.
The actual IO pins are modeled by the bundled pins along the block boundary, 
as described in Section~\ref{sec:autoclustering}.
At the top level (i.e., root cluster of $T_P$), 
there \textcolor{black}{are} no other clusters of the parent level.
Below the top level, i.e., when we are working on the physical clusters at intermediate 
levels of the physical hierarchy, the outlines and locations of clusters of the parent level
have \textcolor{black}{already} been calculated, and behave 
\textcolor{black}{like}
fixed terminals.
For example, in Figure \ref{fig:hier},
when we place clusters $c_{13}$ and $c_{14}$, 
other clusters of the parent level ($c_1$, $c_3$ and $c_8$) behave like fixed terminals.
In our implementation, we assume that the bundled pin of each cluster is at the center 
of the cluster.} We use Sequence Pair~\cite{MurataFNK96} to represent a given (floorplanned) arrangement 
of clusters, and Simulated Annealing~\cite{KirkpatrickGV83} to optimize a heuristic
cost function.
\textcolor{black}{
Further, we adopt ``multi-start'' scheme to improve
the performance of Simulated Annealing and we set
the number of threads to 10 in our experiments.
As applied in this phase, the Sequence Pair-based annealing supports four solution perturbation
(move) operators with respective probabilities 0.3, 0.3, 0.3 and 0.1:
\begin{itemize}[noitemsep,topsep=0pt,leftmargin=*]
    \item \textbf{Op1:} Swap two clusters in the first sequence; 
    \item \textbf{Op2:} Swap two clusters in the second sequence; 
    \item \textbf{Op3:} Swap two clusters in each of both sequences; and
    \item \textbf{Op4:} Resize a cluster. For standard-cell clusters and mixed clusters, 
    we use the same resizing algorithm as in \cite{ChenC06}; 
    for macro clusters, we randomly pick one shape of its shapes.
\end{itemize}
}

\noindent
To improve the QoR of the floorplan,
\textcolor{black}{
we handle the following constraints as {\em RTL-MP} does.}\footnote{
\textcolor{black}{Detailed
algorithms for handling these constraints 
are given in \cite{RTL-MP}.
And the implementation is available in \cite{Hier-RTLMP}.}}
\begin{itemize}[noitemsep,topsep=0pt,leftmargin=*]
    \item \textbf{Fixed outline:} All clusters should be placed within the fixed outline.
    At the top level (root cluster of the physical hierarchy $T_P$), 
    the outline is the block boundary defined in the input .def file.
    Below the top level, for an intermediate level physical cluster, 
    the outline is determined by the shape and location of its parent cluster.
    \item \textbf{Peripheral bias:} \textcolor{black}{To take routing blockages (e.g., due to
    internal routing within macros)
    into account, leaf clusters with macros should be pushed 
    to peripheries of the boundary.
    This simplifies the routing process by reducing congestion in the center region.}

    \item \textbf{Blockage:} Instances including both macros and standard cells should not overlap with 
    placement blockages. Since preplaced macros can be treated as placement blockages, 
    our macro placer can also handle preplaced macros.
    \item \textbf{Guidance:} All clusters should be placed {\em near} specified regions if users provide such constraints. 
    \textcolor{black}{
    The cluster's guidance is determined by the bounding box that encompasses the guidance for all constituent macros.
    We do not consider the guidance for macros or clusters during 
    multilevel autoclustering.}
    \item \textbf{Notch avoidance: } A decent floorplan should avoid 
    {\em dead space} which cannot be used effectively by P\&R tools.
    \item \textbf{Pin access:} Macros should be kept from blocking the access of IO pins. 
\end{itemize}

\noindent
In summary, the final cost function of our macro placer is 
\begin{equation} \label{eq:cost}
\begin{split} 
  cost &= \alpha \times Area + \beta \times WL  + \gamma \times p_{outline} \\
       & + \zeta \times p_{bias} + \eta \times p_{blockage} \\
       &+  \theta \times p_{guidance} 
             + \lambda  \times p_{notch} 
\end{split}
\end{equation}

\noindent
where $Area$ is the area of the current floorplan, $WL$ is the 
wirelength (HPWL), $p_{outline}$ is the penalty for violating the
fixed outline constraint, $p_{bias}$ is the penalty to promote 
macro peripheral bias,
$p_{blockage}$ is the penalty for pin access and macro blockage,
$p_{guidance}$ is the penalty for macro guidance, $p_{notch}$ is the
penalty for notch regions, and $\alpha$, 
$\beta$, $\gamma$, 
$\zeta$, $\eta$, 
$\theta$ and $\lambda$ are the corresponding weights. 
$Area$, $WL$, $p_{outline}$, $p_{bias}$, $p_{blockage}$,  $p_{guidance}$ 
and $p_{notch}$ are all normalized by the corresponding initial 
value. 
\textcolor{black}{
The default values of these weights are available in \cite{Hier-RTLMP},
and the effects of tuning these weights are studied in Section \ref{sec:experiment}.}

\subsection{Placement of Macros}
\label{sec:pinAlignmentEngine}

After the placement of clusters, we know the position 
and shape for each cluster. We next determine the location and 
orientation of macros in each leaf macro cluster, one macro cluster 
at a time. For a given macro cluster {\em A}, 
we extract the connections between macros in {\em A} and in other 
clusters. Here, other clusters behave like fixed terminals. 
We use Sequence Pair~\cite{MurataFNK96} to represent macro placement
in {\em A} and Simulated Annealing~\cite{KirkpatrickGV83} to optimize 
the cost function.
We use four solution perturbation (move) operators in 
the annealing, with respective probabilities 
0.3, 0.3, 0.3 and 0.1:
\begin{itemize}[noitemsep,topsep=0pt,leftmargin=*]
    \item \textbf{Op1:} Swap two macros in the first sequence; 
    \item \textbf{Op2:} Swap two macros in the second sequence; 
    \item \textbf{Op3:} Swap two macros in each of both sequences; and
    \item \textbf{Op4:} Flip all the macros.
\end{itemize}
The cost function used in this step is 
\begin{equation}
\label{eq:cost}
       cost  = \alpha \times Area + \beta \times WL + \gamma \times p_{outline} 
             + \theta \times p_{guidance} 
\end{equation}
where $Area$ is the area of the current macro packing, $WL$ is the 
wirelength (HPWL), $p_{outline}$ is the penalty for violating the
fixed-outline constraint, $p_{guidance}$ is the penalty for macro guidance,
and $\alpha, \beta$, $\gamma$ and $\theta$
are corresponding weights.

\section{EXPERIMENTAL VALIDATION}
\label{sec:experiment}

\textcolor{black}{
{\em Hier-RTLMP} is implemented with approximately 12K lines of C++ 
with a Tcl command line interface on top of the  OpenROAD~\cite{KahngS21, openroad-github} infrastructure.\footnote{
\textcolor{black}{
We make public with permissive open-source license all source code at \cite{Hier-RTLMP}.}}
We have validated our macro placer using multiple designs including Ariane \cite{ariane}, 
BlackParrot (Quad-Core) \cite{bp_quad}, MemPool Group \cite{mempool_group}, Tabla09 \cite{RTML}, 
Tabla01 \cite{RTML} and Arm Cortex-A53 (CA53), in both open NanGate45 (NG45) and commercial GlobalFoundries 12nm (GF12) enablements.
{\em Tabla09} and {\em Tabla01} are machine learning accelerators generated by 
an open-source machine learning hardware generator \cite{RTML}.
We use the {\em bsg\_fakeram} \cite{bsg_fakeram} memory generator to 
generate SRAMs for NanGate45 enablement.
The commercial GlobalFoundries 12nm enablement is a commercial foundry 12nm technology (13 metal
layers) with cell library and memory generators from a leading IP
provider.
Table~\ref{tab:benchmark} summarizes information about our designs.}

\vspace{0.5cm}

\begin{table}[!h]
  \caption{\small \textcolor{black}{Benchmarks.
  The clock periods for testcases implemented in GF12 
  are not specified, to protect foundry IP.
  $Min\_AR$ and $Max\_AR$ respectively denote the minimum and maximum aspect ratios of macros within the testcase.
  {\em Area Ratio} refers to the ratio between the maximum and minimum areas of macros in the testcase.}}
    \resizebox{1.0\columnwidth}{!} {
    \centering
    \begin{tabular}{|c|c|c|c|c|c|c|S|}
    \hline
    Designs & Enablements &  Macros 
            & {\makecell{ Std \\ Cells }}
            & {\makecell{ Clock \\ Period }}
            & Utilization 
            & {\makecell{ \textcolor{black}{ Min\_AR} \\  \textcolor{black}{Max\_AR }}}
            & {\makecell{ \textcolor{black}{ Area } \\ \textcolor{black}{ Ratio }}} \\ \hline
    Ariane  & NG45 & 133 & 118K & 1.3 ns & 0.70 &  \textcolor{black}{ 1.53 / 2.31 } &  \textcolor{black}{ \:  2.37 } \\ \hline
    BlackParrot & NG45 & 220 & 769K & 5.2 ns & 0.69 &  \textcolor{black}{ 0.61 / 3.24 } &  \textcolor{black}{ 14.14 } \\ \hline
    CA53 & GF12 & 25 & 445K & ----- & 0.73  &  \textcolor{black}{ 0.90 / 4.80 } & \textcolor{black}{ \: 8.83 } \\ \hline
    Ariane & GF12 & 133 & 95K & ----- & 0.68  & \textcolor{black}{ 0.95 / 0.95 } & \textcolor{black}{ \: 1.00 }  \\ \hline
    BlackParrot & GF12 & 196 & 827K & ----- & 0.75 & \textcolor{black}{ 1.38 / 4.54 } & \textcolor{black}{ \: 3.96 }   \\ \hline
    Tabla09 & GF12 & 368 & 245K & -----  & 0.75 & \textcolor{black}{ 0.88 / 3.38 } & \textcolor{black}{ 11.74  } \\ \hline
    Tabla01 & GF12 & 760 & 431K  & ----- & 0.75  & \textcolor{black}{ 1.55 / 3.38 } & \textcolor{black}{ 18.52 } \\ \hline
    MemPool & GF12 & 326 & 2529K  & ----- & 0.62  & \textcolor{black}{ 1.35 / 3.77 } & \textcolor{black}{ \: 2.79 } \\ \hline
    \end{tabular}
    }
    \label{tab:benchmark}
\end{table}

To show the effectiveness of our macro placer, the following 
three scenarios are evaluated and compared.\footnote{\cite{LinDYCL21} 
has reported an excellent dataflow-driven macro placer.
Unfortunately, no testcases or executables can be released by 
their group. Our previous work \cite{RTL-MP} also tried to compare 
against the original mixed-size placer  ({\em TritonMacroPlacer}, 
or {\em tmp}) in OpenROAD, but the tool was unable to generate legal 
floorplans for many of our designs.}
\begin{itemize}[noitemsep,topsep=0pt,leftmargin=*]
    \item {\em Comm}: Macro placement is performed using a 2021 release of 
    a state-of-the-art commercial P\&R tool (unnamed due to EULA 
    restrictions) with its latest macro placement option.
    \item {\em RTL-MP}: Macros are placed by our previous work, {\em RTL-MP}~\cite{RTL-MP}.
    \item \textcolor{black}{{\em Hier-RTLMP}: Results are obtained using our new macro placer.}
\end{itemize}
Our experiments use the following flow. 
(1) We first synthesize the design using a state-of-the-art commercial synthesis 
tool, preserving the logical hierarchy. 
\textcolor{black}{(2) Next, we determine the core size of the testcase, and place all the IO pins according to the determined core size using a manually-developed script.} 
The utilization for each testcase is presented in Table \ref{tab:benchmark}.
(3) Then, the macros are placed using different methods ({\em Comm}, {\em RTL-MP} and {\em Hier-RTLMP}).
\textcolor{black}{(4) Last, all standard cells are placed, and 
all nets are routed, using Cadence Innovus v21.1.} 
Our scripts for power delivery network generation and standard-cell placement and routing are similar 
to those publicly visible in the {\em MacroPlacement} GitHub repository~\cite{MacroPlacement-github}.

\subsection{\textcolor{black}{Comparison of {\em Hier-RTLMP} with Other Macro Placers}}
\label{sec:metrics}

\textcolor{black}{
Table \ref{tab:rtlmp_result} and Table \ref{tab:hier_rtlmp_result} show}
\textcolor{black}{
the experiment results 
after completion of post-routing optimization (postRouteOpt) 
\textcolor{black}{starting from different macro placers' solutions on our testcases.}
\textcolor{black}{All the macro placers
are executed using their default parameter settings.
The effect of parameter tuning is discussed in Section \ref{sec:autotune}.}
Rows represent designs, enablements and macro placement flows\textcolor{black}{;} 
columns give metrics \textcolor{black}{of} number of standard cells, 
total routed wirelength, 
power, worst negative slack (WNS), total negative slack (TNS) 
and turnaround time for a single run.
The metrics in GF12 are normalized to protect foundry IP:
(i) standard-cell area is normalized to core area;
(ii) wirelength and power are normalized to the {\em Comm} result;
and (iii) timing metrics (WNS, TNS) are normalized to the clock period which we leave unspecified.}
\textcolor{black}{
In all experiments that we report,
we allow {\em Hier-RTLMP} 
to sweep the target dead space $t\_dead\_space$ from 0.05 to 1.0 with step 0.05, and target utilization $util$ from 0.25 to 1.0 with step 0.1.}
\textcolor{black}{
To reduce the runtime of $t\_dead\_space$
and $util$ sweeping, we set the number of threads to 10.}



As noted in Section \ref{sec:shape_function}, the parameter sweep stops with the
first valid placement of clusters (in practice, within 10 to 30 trials).
The turnaround time for a single run reported includes the total runtime of all
the trials (see Section \ref{sec:runtime} for detailed runtime analysis).\footnote{\textcolor{black}{A single run means running the corresponding macro placer once, without any parameter sweeping or autotuning.}}
{\em Hier-RTLMP} outperforms \textcolor{black}{the} commercial macro placer and {\em RTL-MP} for almost all the testcases.

\begin{figure}[!htb]
     \centering
    \begin{subfigure}[b]{0.22\textwidth}
         \centering
         \includegraphics[width=\textwidth]{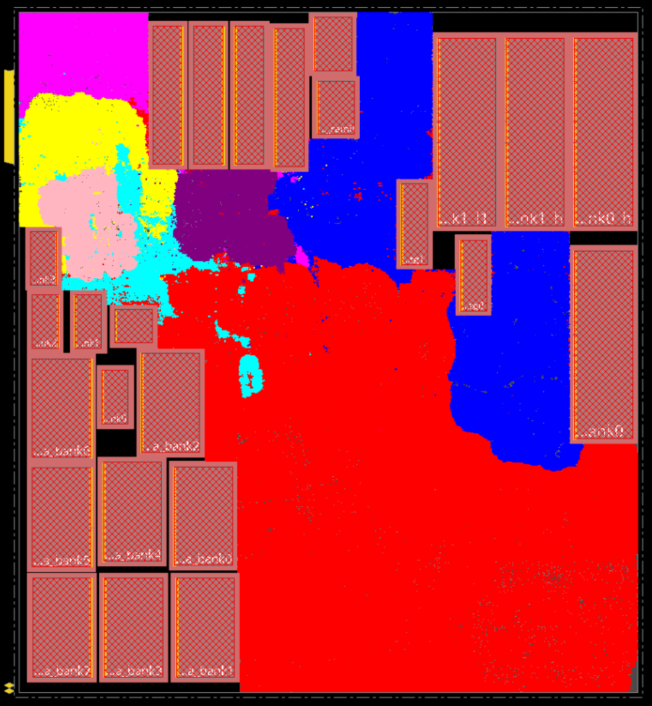}
         \caption{\textcolor{black}{CA53 (GF12) / Comm}}
    \end{subfigure}
    \begin{subfigure}[b]{0.22\textwidth}
         \centering
         \includegraphics[width=\textwidth]{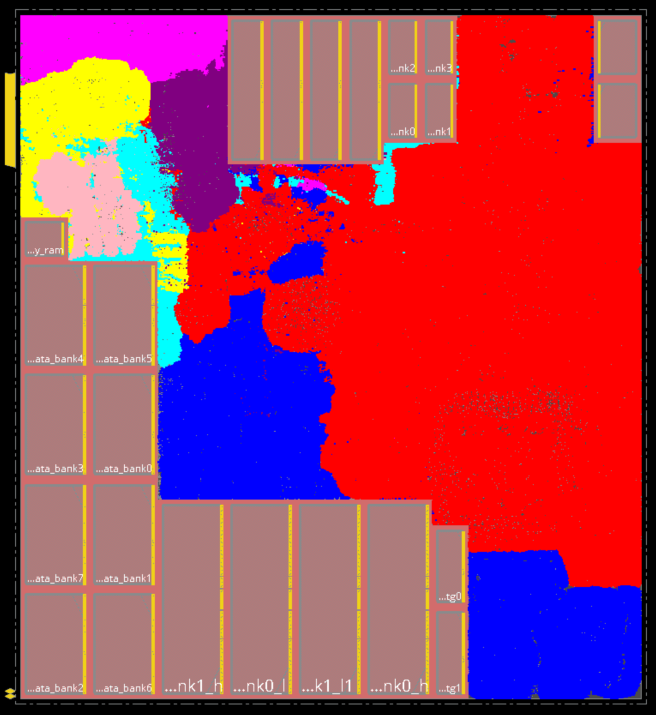}
         \caption{\textcolor{black}{CA53 (GF12) / RTL-MP}}
     \end{subfigure}
     \begin{subfigure}[b]{0.22\textwidth}
         \centering
         \includegraphics[width=\textwidth]{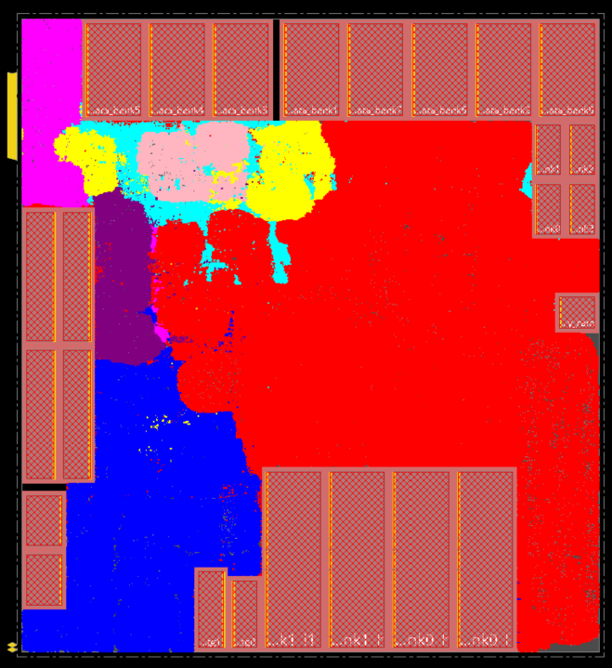}
         \caption{\textcolor{black}{CA53 (GF12) / Hier-RTLMP}}
    \end{subfigure}
    \begin{subfigure}[b]{0.22\textwidth}
         \centering
         \includegraphics[width=\textwidth]{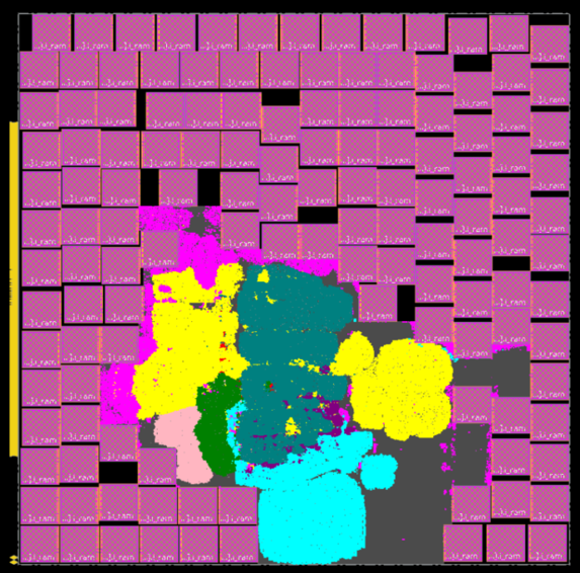}
         \caption{\textcolor{black}{Ariane (GF12) / Comm}}
    \end{subfigure}
     \begin{subfigure}[b]{0.22\textwidth}
         \centering
         \includegraphics[width=\textwidth]{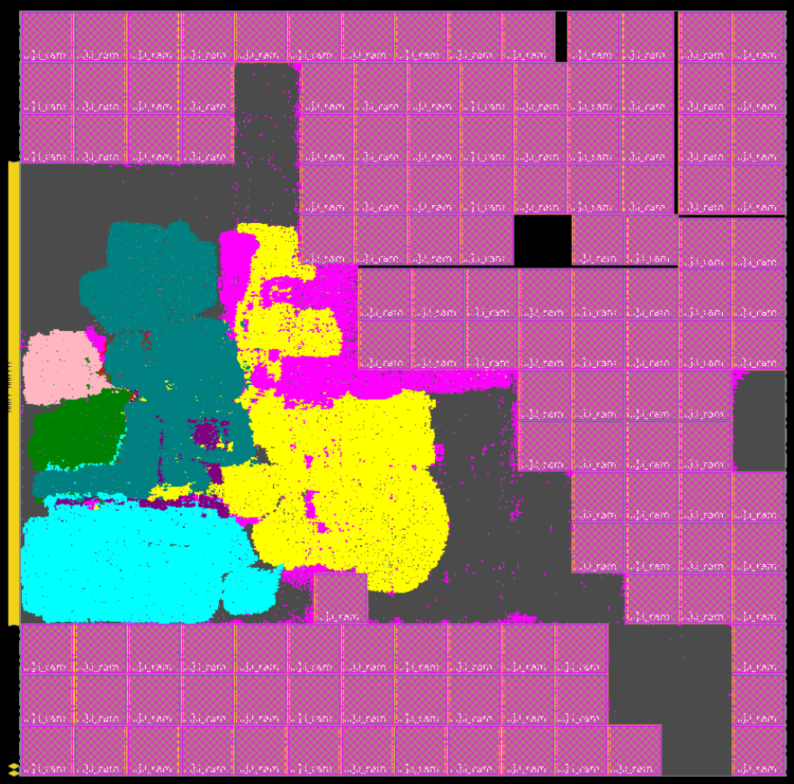}
         \caption{\textcolor{black}{Ariane (GF12) / RTL-MP}}
    \end{subfigure}
    \begin{subfigure}[b]{0.22\textwidth}
         \centering
         \includegraphics[width=\textwidth]{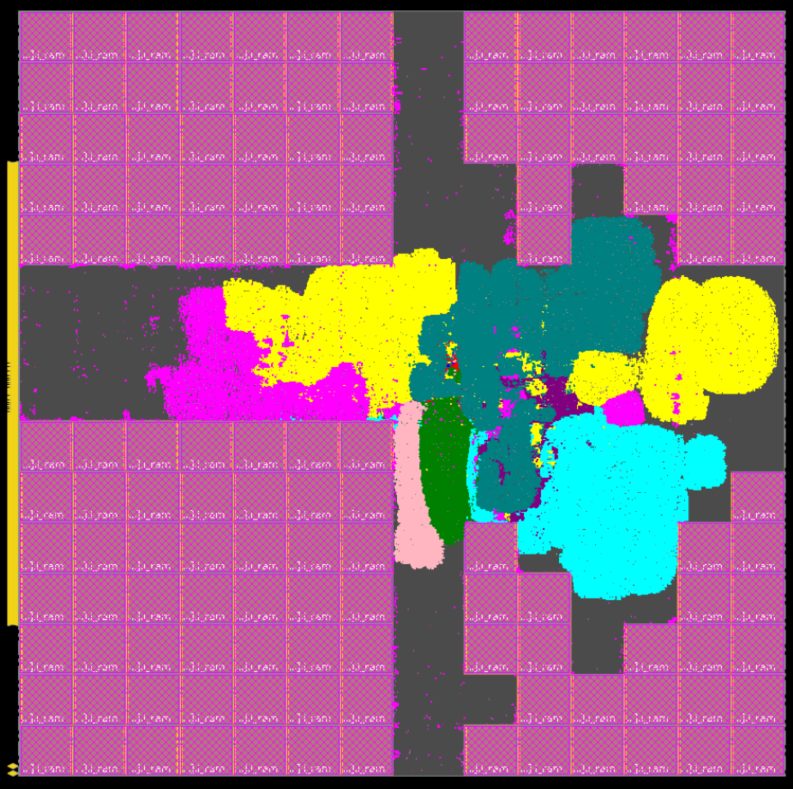}
         \caption{\textcolor{black}{Ariane (GF12) / Hier-RTLMP}}
     \end{subfigure}
    \centering
    \caption{
    \textcolor{black}{
    Macro placements generated by {\em Comm}, {\em RTL-MP} and {\em Hier-RTLMP}.}}
    \label{fig:HierRTLMPvsRTLMP}
\end{figure}

\noindent
\textbf{Compared to {\em RTL-MP}.} 
\textcolor{black}{
Table \ref{tab:rtlmp_result} shows that 
{\em Hier-RTLMP} reduces runtime compared to {\em RTL-MP} by at least $13\times$.
For CA53 (GF12), {\em Hier-RTLMP} generates worse results than {\em RTL-MP}.
\textcolor{black}{However,} {\em Hier-RTLMP} can achieve similar TNS through applying autotuning enhancement (see Section \ref{sec:autotune}).
For Ariane (GF12), {\em Hier-RTLMP} generates better results than {\em RTL-MP}.}
\textcolor{black}{
The corresponding layouts of CA53 (GF12) and Ariane (GF12) are presented in Figure \ref{fig:HierRTLMPvsRTLMP}.}

\vspace{0.1cm}
\begin{table}[!h]
\caption{\textcolor{black}{
Metrics of {\em Comm}, {\em RTL-MP} and {\em Hier-RTLMP}
for designs in commercial GF12 enablements.
We highlight best values of \textcolor{black}{timing} metrics in blue bold font.
Data points for GF12 are normalized.
}
}

\label{tab:rtlmp_result}
\resizebox{1\columnwidth}{!} {
\centering
\begin{tabular}{|c|c|c|c|c|c|c|c|}
\hline \Xhline{3\arrayrulewidth}
\multicolumn{1}{|l|}{\makecell{ Design\\ (Enablement) }}             
    & \makecell{ Macro\\Placer}
    & \makecell{Std Cell \\ Area ($mm^2$)}   & \makecell{WL \\($m$)}   &  \makecell{Power \\($mW$)}
    & \makecell{ \textcolor{black}{WNS} \\ \textcolor{black}{(ps)} } 
    & \makecell{ \textcolor{black}{TNS} \\ \textcolor{black}{(ns)} }
    & \makecell{ TAT\\($min$)}
    \\ \hline \Xhline{3\arrayrulewidth}
    
\multirow{3}{*}{\makecell{CA53\\ (GF12) }}                                                  
        & {\em Comm}         
        & 0.166
        & 1.00 
        & 1.00
        & -0.23  
        & -277
        & 68 \\  \cline{2-8}
        & {\em RTL-MP}
        & 0.167  & 1.04  & 1.01  
        & \textbf{\textcolor{blue}{-0.23}}  
        & \textbf{\textcolor{blue}{-145}}   
        & 71 \\  \cline{2-8} 
        & {\em Hier-RTLMP}     
        & 0.168  
        & 1.08
        & 1.03      
        & -0.26   
        & -167
        & 5 \\ \hline \Xhline{3\arrayrulewidth}

\multirow{3}{*}{\makecell{Ariane\\(GF12)}}                           
        & {\em Comm}         
        & 0.045
        & 1.00  
        & 1.00
        & -0.17
        & -156
        & 11 \\  \cline{2-8}
        & {\em RTL-MP}
        & 0.045  & 1.29   & 1.02
        & -0.13 & -108   & 200 \\  \cline{2-8}
        & {\em Hier-RTLMP}     
        & 0.045   & 1.21  & 1.01          
        & \textbf{\textcolor{blue}{-0.09}} 
        & \textbf{\textcolor{blue}{-63}}  
        & 9 \\ \hline \Xhline{3\arrayrulewidth}
\end{tabular}
}
\end{table}

\vspace{0.5cm}

\begin{figure*}
    \centering
   \begin{subfigure}[b]{0.225\textwidth}
         \centering
         \includegraphics[width=\textwidth]{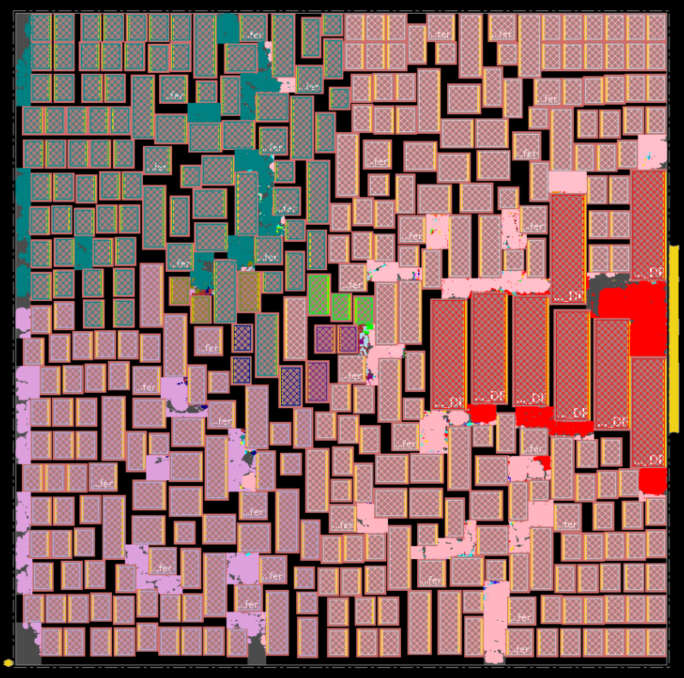}
         \caption{\textcolor{black}{Tabla09 (GF12) / Comm}}
    \end{subfigure}
    \begin{subfigure}[b]{0.23\textwidth}
         \centering
         \includegraphics[width=\textwidth]{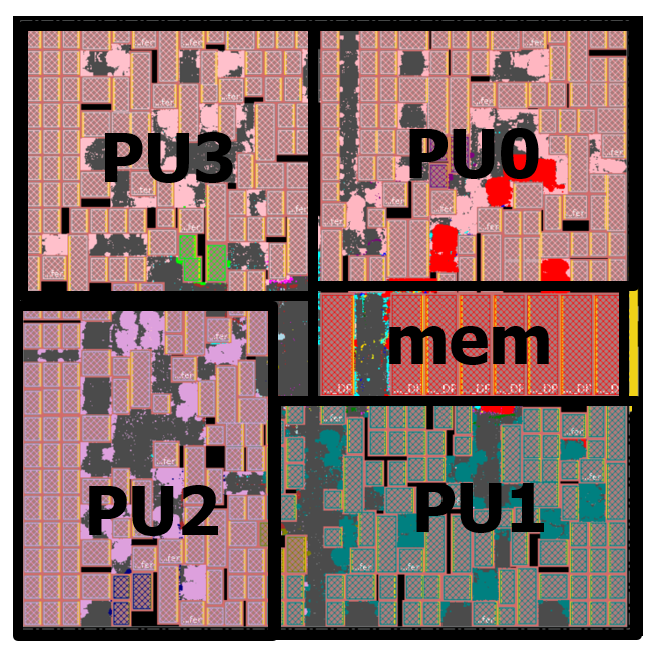}
         \caption{\textcolor{black}{Tabla09 (GF12) / Hier-RTLMP}}
     \end{subfigure}
     \begin{subfigure}[b]{0.24\textwidth}
         \centering
         \includegraphics[width=\textwidth]{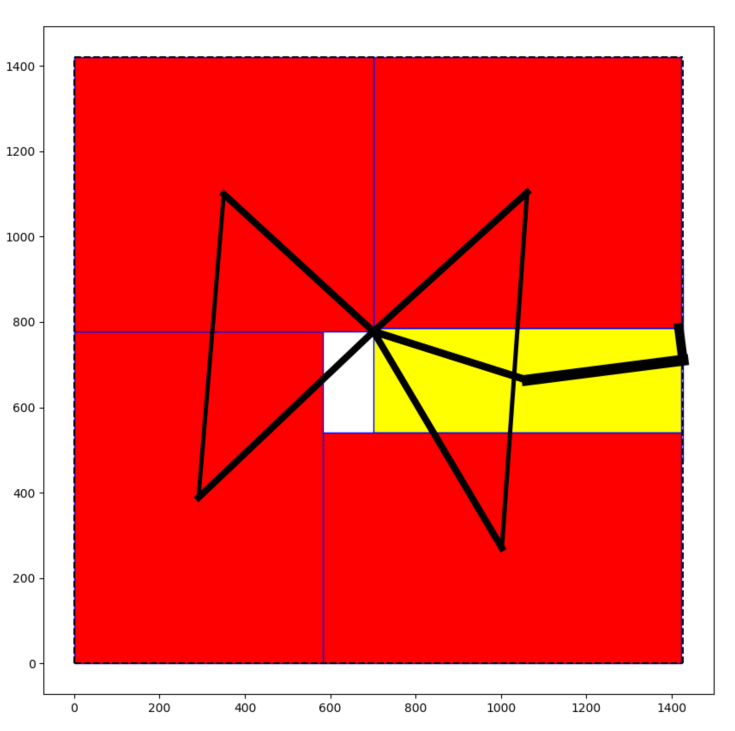}
         \caption{\textcolor{black}{Tabla09 (GF12) dataflow}}
    \end{subfigure}
    \begin{subfigure}[b]{0.24\textwidth}
         \centering
         \includegraphics[width=\textwidth]{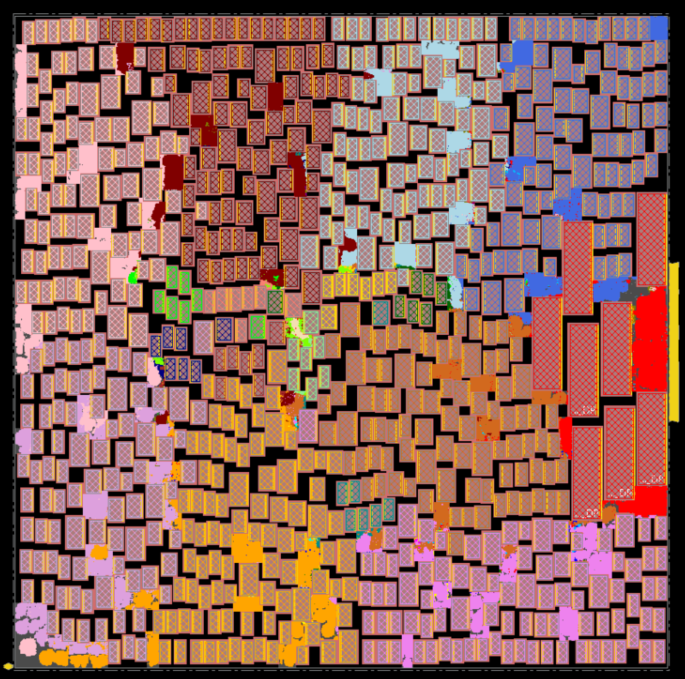}
         \caption{\textcolor{black}{Tabla01 (GF12) / Comm}}
    \end{subfigure}
     \begin{subfigure}[b]{0.24\textwidth}
         \centering
         \includegraphics[width=\textwidth]{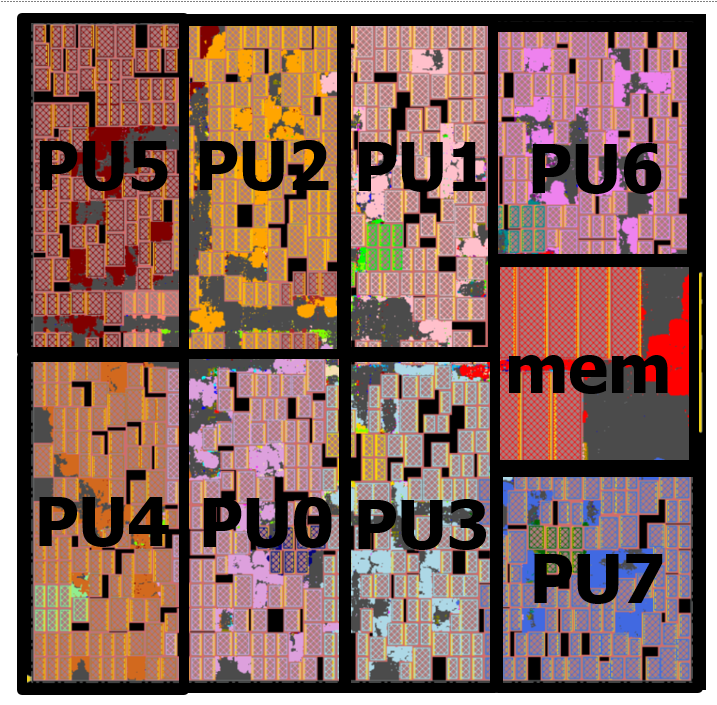}
         \caption{\textcolor{black}{Tabla01 (GF12) / Hier-RTLMP}}
     \end{subfigure}
     \begin{subfigure}[b]{0.24\textwidth}
         \centering
         \includegraphics[width=\textwidth]{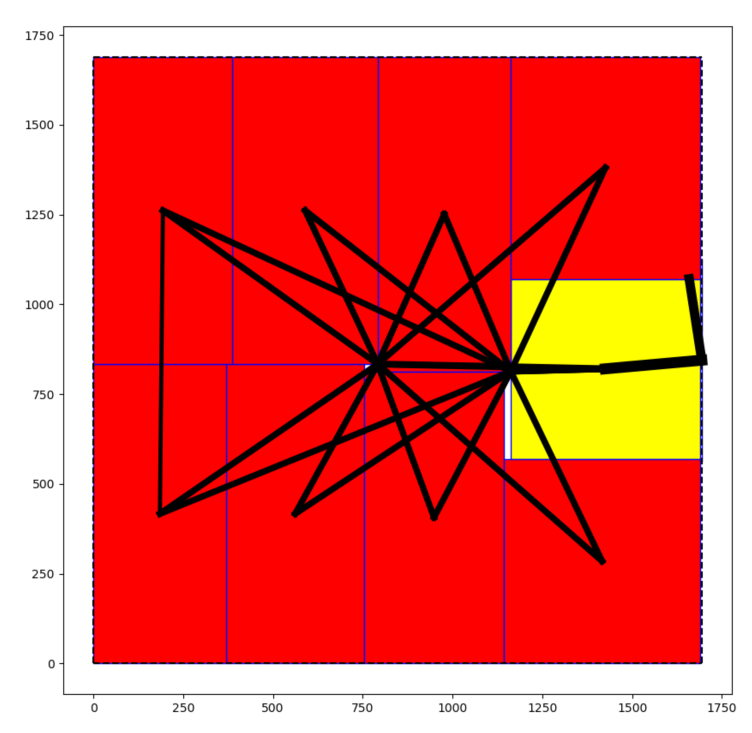}
         \caption{\textcolor{black}{Tabla01 (GF12) dataflow}}
    \end{subfigure}
    \begin{subfigure}[b]{0.24\textwidth}
         \centering
         \includegraphics[width=\textwidth]{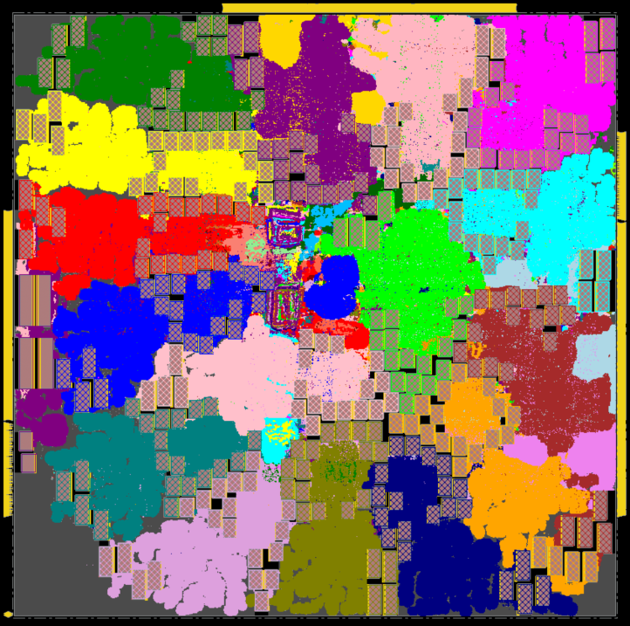}
         \caption{\textcolor{black}{MemPool (GF12) / Comm}}
    \end{subfigure}
     \begin{subfigure}[b]{0.24\textwidth}
         \centering
         \includegraphics[width=\textwidth]{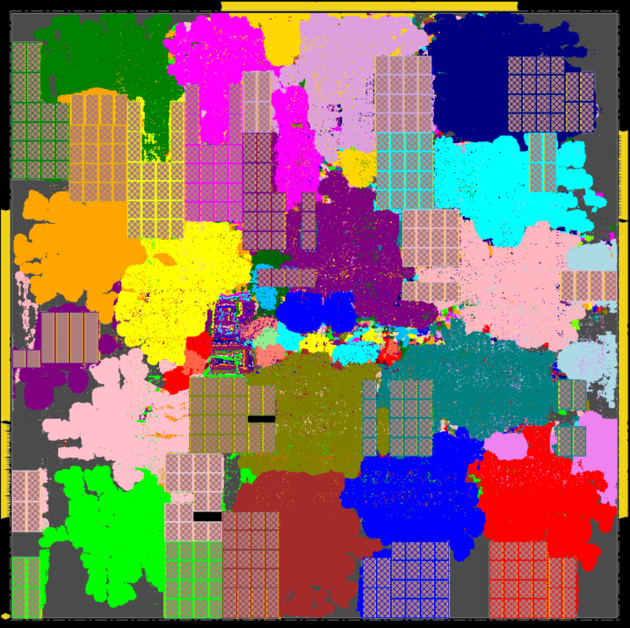}
         \caption{\textcolor{black}{MemPool (GF12) / Hier-RTLMP}}
     \end{subfigure}    \centering
    \caption{
    \textcolor{black}{
    Macro placements generated by {\em Comm} and {\em Hier-RTLMP}.}
    }
    \label{fig:HierRTLMPvsCMP}
\end{figure*}

\noindent
\textbf{Compared to commercial macro placers.} 
\textcolor{black}{
Table \ref{tab:hier_rtlmp_result} shows that {\em Hier-RTLMP} achieves 
much better timing in terms of WNS and TNS within similar or less runtime.\footnote{
\textcolor{black}{
{\em RTL-MP} fails on Table \ref{tab:hier_rtlmp_result} testcases for two reasons: (i) its database is not fully compatible
with the NanGate45 technology node, and (ii) its 
\textcolor{black}{capability is} limited to single-level autoclustering (see Section \ref{sec:autoclustering}), which prevents it from handling testcases with hundreds of macros.
}}}
\textcolor{black}{
Besides, {\em Hier-RTLMP} can identify the dataflow of design and place macros following the dataflow.
For example, for the Tabla09 design,
Figure \ref{fig:HierRTLMPvsCMP}(b) shows the layout for the macro placement generated by 
{\em Hier-RTLMP} and Figure \ref{fig:HierRTLMPvsCMP}(c) shows 
the placement for the \textcolor{black}{child} clusters of the root (top-level) cluster.
At the root level of the physical hierarchy, five clusters are 
mixed clusters containing both macros and standard cells. 
There is one IO cluster containing memories ($mem$, yellow rectangle), 
and four functional units each of which is an individual mixed cluster ($PU0$ to $PU3$, red rectangles).
The standard-cell clusters at the top level,
which are ``tiny clusters'' (Section \ref{sec:shape_function}),
contain muxing logic that processes the IOs and interfaces with the four functional units.
As can be seen from Figure \ref{fig:HierRTLMPvsCMP}(b),
the placement follows the dataflow with the IO cluster close to the IOs and 
the  standard-cell cluster in the middle of the four functional unit clusters. 
The black lines \textcolor{black}{in Figure \ref{fig:HierRTLMPvsCMP}(c) show the bundled net connections.
The Tabla01 design has a similar architecture as Tabla09, 
but with eight functional units ($PU0$ to $PU7$).
The results are presented in Figure \ref{fig:HierRTLMPvsCMP}(d)-(f).}}
\textcolor{black}{
The layouts for Mempool (GF12) are presented in Figure \ref{fig:HierRTLMPvsCMP}(g)-(h).}

\begin{table}
\caption{\textcolor{black}{Metrics of {\em Comm} and {\em Hier-RTLMP} for designs in
open NG45 and commercial GF12 enablements.
We highlight best values of \textcolor{black}{timing} metrics in blue bold font.
Data points for GF12 are normalized.
}
}
\label{tab:hier_rtlmp_result}
\resizebox{1\columnwidth}{!} {
\centering
\begin{tabular}{|c|c|c|c|c|c|c|c|}
\hline \Xhline{3\arrayrulewidth}
\multicolumn{1}{|l|}{\makecell{ Design\\ (Enablement) }}             
    & \makecell{ Macro\\Placer}
    & \makecell{Std Cell \\ Area ($mm^2$)}   & \makecell{WL \\($m$)}   &  \makecell{Power \\($mW$)}
    & \makecell{ WNS\\($ps$)} &\makecell{TNS \\($ns$) }
    & \makecell{ TAT\\($min$)}
    \\ \hline \Xhline{2\arrayrulewidth}
    
\multirow{2}{*}{\makecell{Ariane\\ (NG45) }}                                                       
        & {\em Comm}         
        & 0.247 & 4.35 & 835 & -258 & -629 & 4 \\  \cline{2-8}
        & {\em Hier-RTLMP}     
        & 0.247  & 5.27  & 833            
        & \textbf{\textcolor{blue}{-97}}   
        & \textbf{\textcolor{blue}{-55}}  
        & 8 \\ \hline \Xhline{3\arrayrulewidth}

\multirow{2}{*}{\makecell{BlackParrot\\ (NG45) }}                                   
        & {\em Comm}         
        & 1.926    
        & 24.62
        & 4461
        & -193    
        & -2113     
        & 23 \\  \cline{2-8}
        & {\em Hier-RTLMP}     
        & 1.924 & 28.73 & 4509           
        & \textbf{\textcolor{blue}{-100}}    
        & \textbf{\textcolor{blue}{-170}}
        & 28 \\ \hline \Xhline{3\arrayrulewidth}
        
\multirow{2}{*}{\makecell{BlackParrot\\ (GF12) }}                    
        & {\em Comm}         
        & 0.21 
        & 1.00
        & 1.00
        & -0.10
        & -1147
        & 118 \\  \cline{2-8}
        & {\em Hier-RTLMP}     
        & 0.21   & 1.20  & 1.02           
        & \textbf{\textcolor{blue}{-0.08}}    
        & \textbf{\textcolor{blue}{-444}}  
        & 69 \\ \hline \Xhline{3\arrayrulewidth}
        
\multirow{2}{*}{\makecell{Tabla09 \\ (GF12) }}                       
        & {\em Comm}         
        & 0.05
        & 1.00   
        & 1.00
        & -0.52   
        & -180
        & 36 \\  \cline{2-8}
        & {\em Hier-RTLMP}     
        & 0.05  
        & 0.89   
        & 0.95        
        & \textbf{\textcolor{blue}{-0.18}}    
        & \textbf{\textcolor{blue}{-74}}   
        & 37 \\ \hline \Xhline{3\arrayrulewidth}

\multirow{2}{*}{\makecell{Tabla01 \\ (GF12) }}                                
        & {\em Comm}         
        & 0.06 
        & 1.00  
        & 1.00
        & -0.51  
        & -125
        & 40 \\  \cline{2-8}
        & {\em Hier-RTLMP}     
        & 0.06  
        & 0.92
        & 0.93        
        & \textbf{\textcolor{blue}{-0.04}}  
        & \textbf{\textcolor{blue}{0}}  
        & 105 \\ \hline \Xhline{3\arrayrulewidth}

\multirow{2}{*}{\makecell{MemPool \\ (GF12) }}                                
        & {\em Comm}         
        & 0.34
        & 1.00   
        & 1.00
        & -0.17  
        & -1305
        & 554 \\  \cline{2-8}
        & {\em Hier-RTLMP}     
        & 0.34 
        & 1.12  
        & 1.05         
        & \textcolor{blue}{\textbf{-0.16}} 
        & \textcolor{blue}{\textbf{-1227}}
        & 167 \\ \hline \Xhline{3\arrayrulewidth}
\end{tabular}
}
\end{table}

\subsection{\textcolor{black}{Effect of Timing Awareness}}

\begin{figure}[!h]
    \centering
    \includegraphics[width=0.9\columnwidth]{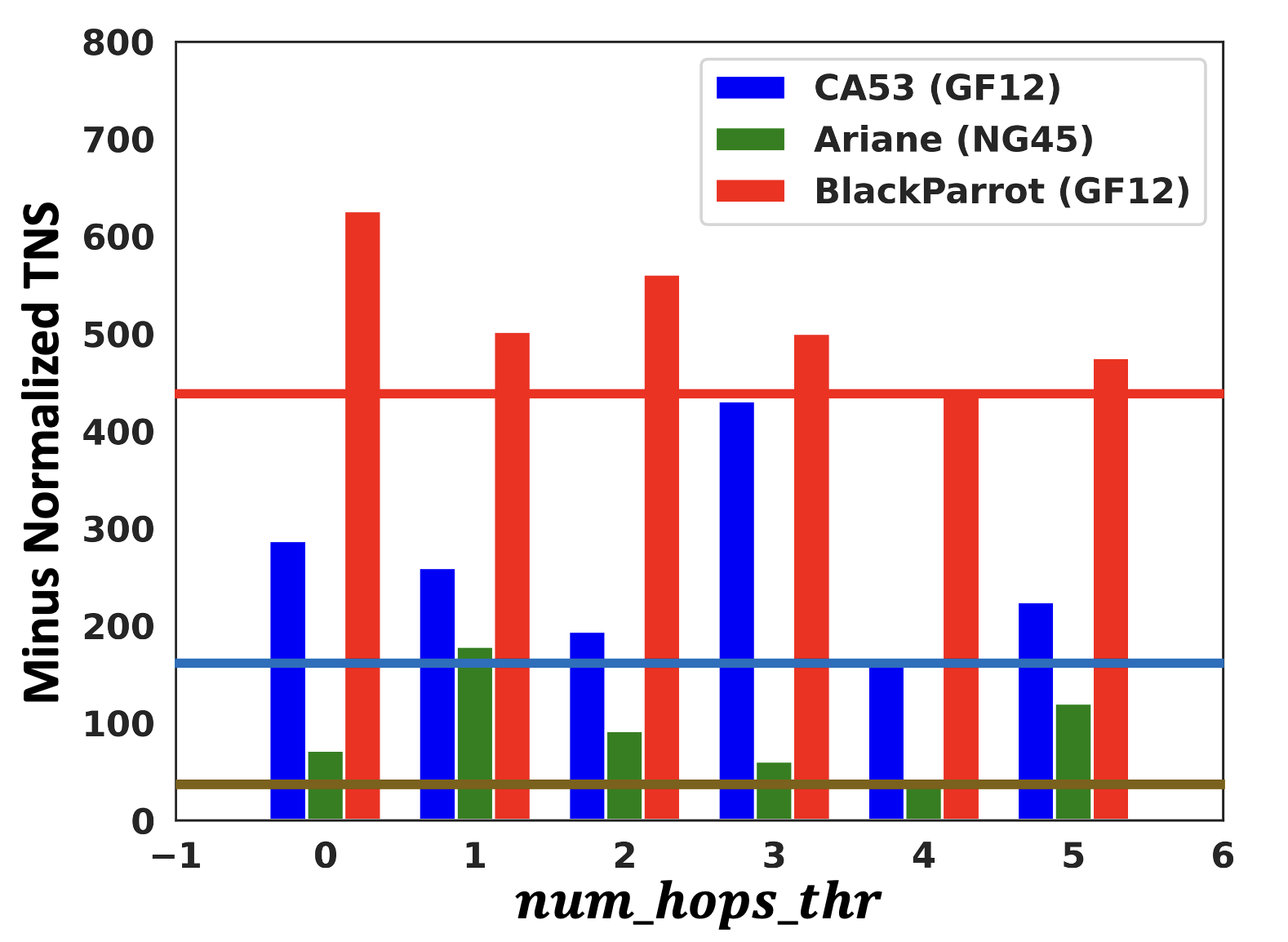}
    \caption{\textcolor{black}{The effect of
    the threshold $num\_hops\_thr$ 
    applied for timing closure in Equation (\ref{eq:num_hops}).}}
    \label{fig:num_hops}
\end{figure}

\textcolor{black}{
{\em Hier-RTLMP} captures the multiple stages of timing paths between clusters through
adding virtual connections between clusters (Section \ref{subsec:multilevel-autoclustering}).
Figure \ref{fig:num_hops} shows the effect of 
varying the threshold of
$num\_hops$ (the length of the shortest path of registers between paths) in Equation (\ref{eq:num_hops}),
which affects the computation of timing-related virtual connections between clusters.
Here we use Ariane (NG45), CA53 (GF12) and BlackParrot (GF12) as our testcases, and sweep 
$num\_hops\_thr$ across values \{0, 1, 2, 3, 4, 5\},
where $num\_hops\_thr = 0$ means that no multiple stages of timing paths are considered. 
We can see that $num\_hops\_thr = 4$ (default value) gives consistently good results
in terms of postRouteOpt TNS.}

\subsection{\textcolor{black}{Ablation Study and Autotuning Enhancement}}
\label{sec:autotune}

\textcolor{black}{
We now discuss the effect of tuning parameters on {\em Hier-RTLMP}.
There are two common approaches for tuning parameters:
(i) ablation study, i.e., sweeping the value of one parameter, 
the remaining parameters being fixed at their default values;
and (ii) autotuning, i.e., applying optimization algorithms like
Bayesian optimization \cite{tune} to tune multiple parameters simultaneously and 
figure out the ``optimal'' parameter combination. 
In this work, we study both approaches and show the possible
PPA (power, performance and area) improvement contributed by parameter tuning.
In the following parameter tuning experiments, 
we use Ariane (NG45) and CA53 (GF12) as our testcases,
and tune the parameters $num\_segment$ $(\phi)$ (Section \ref{subsec:multilevel-autoclustering}) and 
$boundary\_weight$ $(\zeta)$ (Equation (\ref{eq:cost})).
The range of  $num\_segment$ $(\phi)$ is [1, 10],
and the range of $boundary\_weight$ $(\zeta)$ is [0, 50).
Both $num\_segment$ and $boundary\_weight$ are integers, thus there are possible 500 
configurations in the search space.\footnote{
\textcolor{black}{
\textcolor{black}{Strictly speaking}, 
$boundary\_weight$ $(\zeta)$ can be any non-negative real number.}}}

\noindent
\textcolor{black}{
\textbf{Ablation study.}
We sweep $num\_segment$ $(\phi)$ values $\{ 1, 2, 3, 4, 5, 6, 7, 8, 9, 10\}$
and $boundary\_weight$ $(\zeta)$ values $\{ 0, 5, 10, 15, 20, 25, 30, 35, 40, 45\}$.
The best values of $\phi$ and $\zeta$, noted as $\phi^*$ and $\zeta^*$ respectively,
are shown in Table \ref{tab:ablation}.}

\begin{figure}
     \centering
     \begin{subfigure}[b]{0.24\textwidth}
         \centering
         \includegraphics[width=\textwidth]{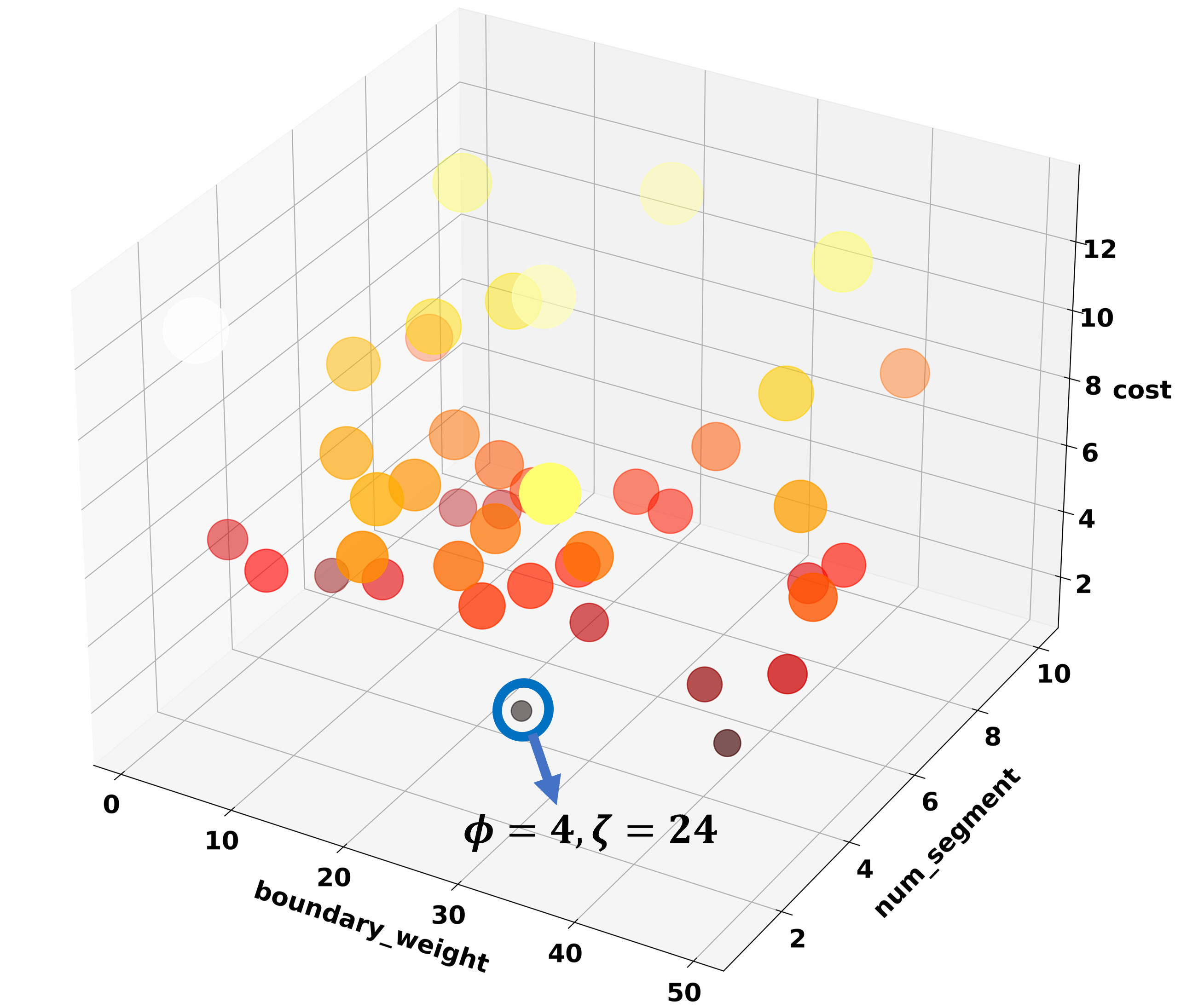}
         \caption{}
     \end{subfigure}
     \hfill
     \begin{subfigure}[b]{0.24\textwidth}
         \centering
         \includegraphics[width=\textwidth]{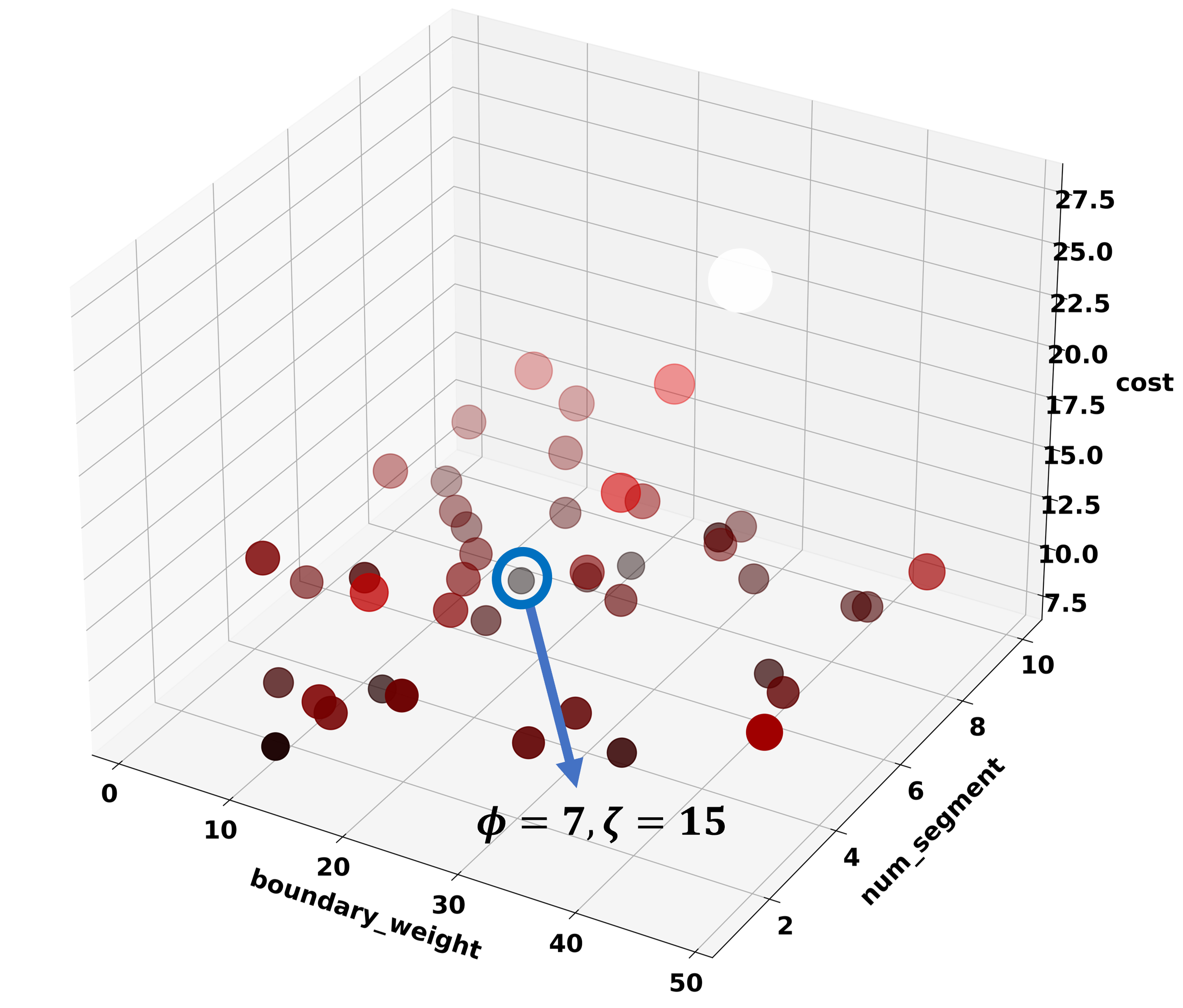}
         \caption{}
    \end{subfigure}
     \caption{\textcolor{black}{Design space explored by the $autotune_{50}$.}
     \textcolor{black}{
     (a) shows the design space explored for Ariane (NG45).
     The best configuration (num\_segment $\phi$ = 4, 
     boundary\_weight $\zeta$ = 24) is highlighted in blue circle;
     and (b) shows the design space explored for CA53 (GF12).
     The best configuration (num\_segment $\phi$ = 7, 
     boundary\_weight $\zeta$ = 15) is highlighted in blue circle.
     }}
    \label{fig:autotune}
\end{figure}

\begin{figure}
     \centering
     \begin{subfigure}[b]{0.225\textwidth}
         \centering
         \includegraphics[width=\textwidth]{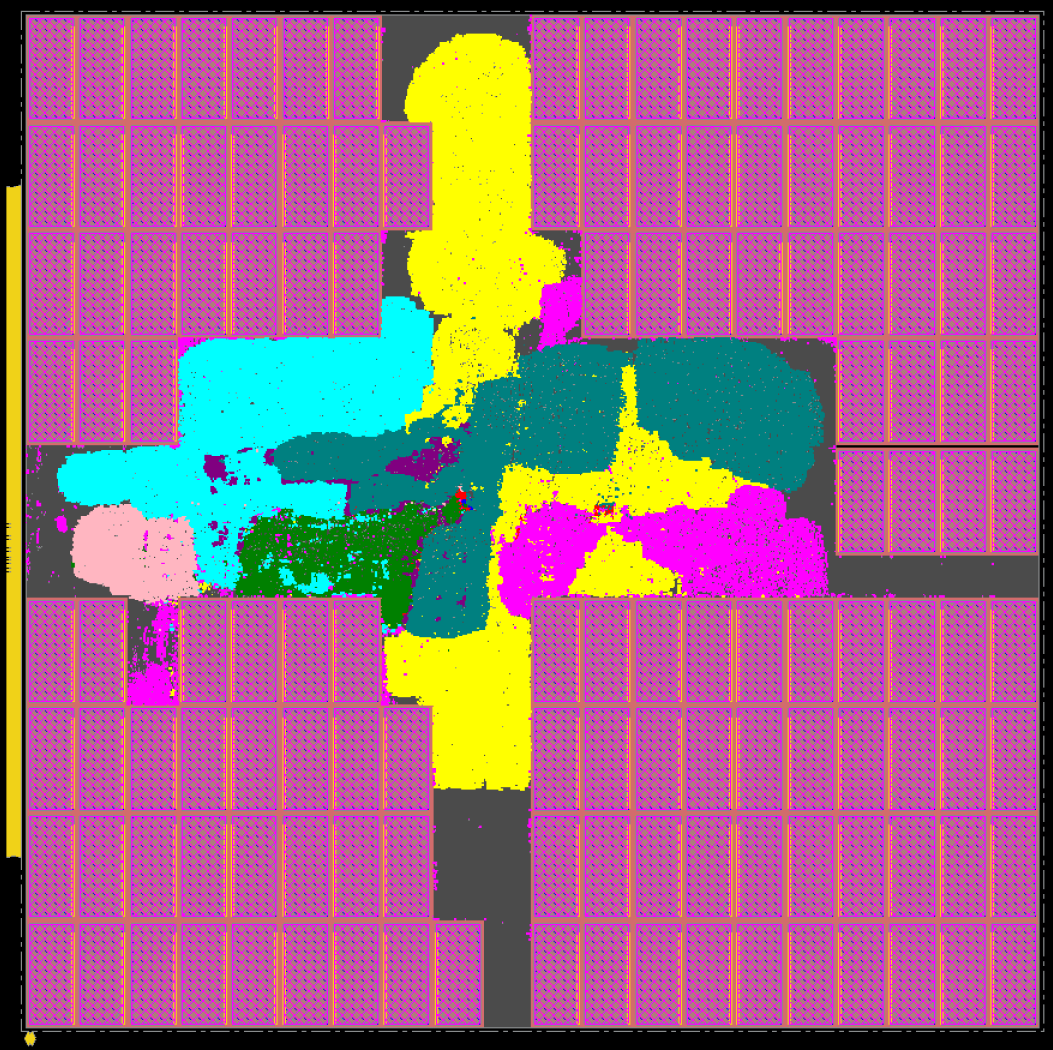}
         \caption{}
     \end{subfigure}
     \hfill
     \begin{subfigure}[b]{0.207\textwidth}
         \centering
         \includegraphics[width=\textwidth]{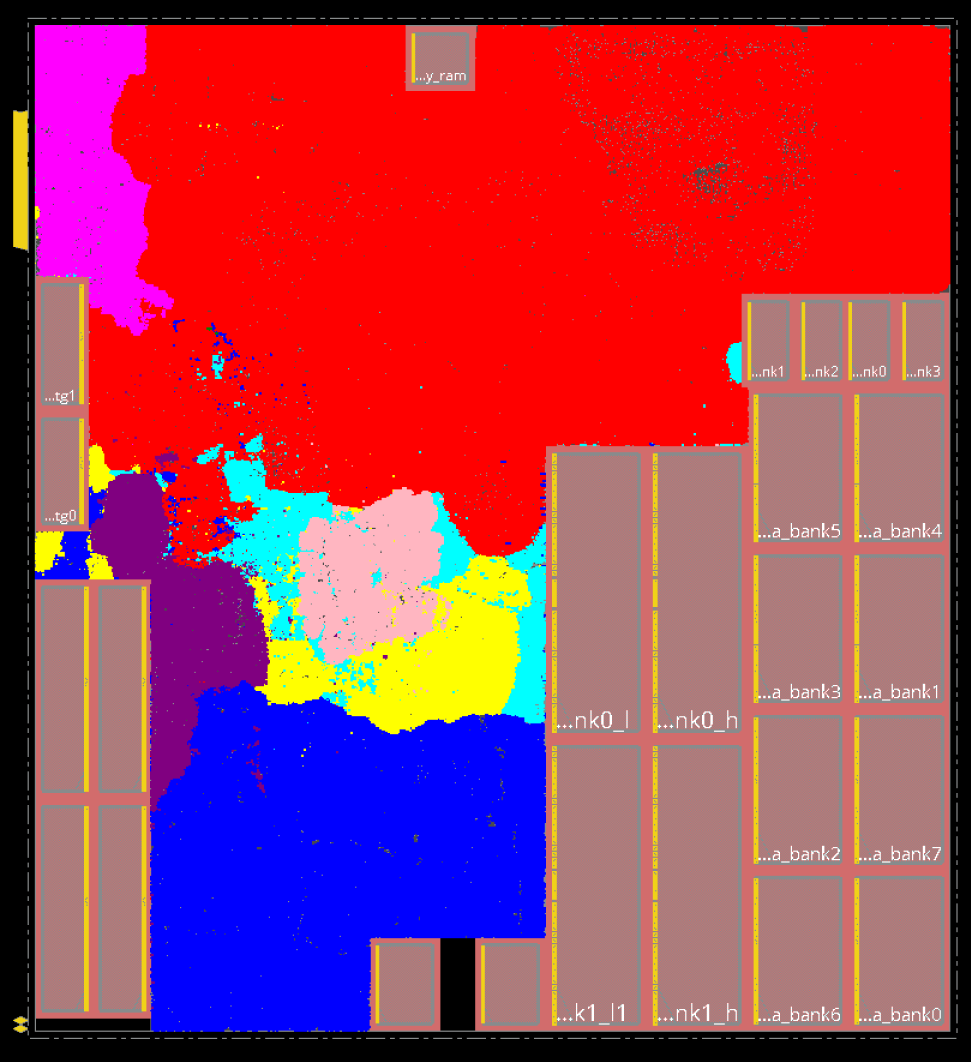}
         \caption{}
    \end{subfigure}
    \caption{
    \textcolor{black}{
    postRouteOpt layouts corresponding to $autotune_{50}^*$, i.e., the best
    results from autotuning with 50 trials. (a) Ariane (NG45) and (b) CA53 (GF12).}}
    \label{fig:autotune_layout} 
\end{figure}

\noindent
\textcolor{black}{
\textbf{Autotuning.} 
We apply the hyperparameter tuning tool Tune \cite{tune} to autotune parameters
$\phi$ and $\zeta$.
An appropriate loss function is needed to guide the search process \cite{tune}.
In our use of Tune, we define the loss function as}
\vspace{-0.05cm}
\begin{equation}
    \centering
    WL\_norm   = \frac{1}{|nets|}\sum_{net}{\frac{length(net)}{core.width + core.height}} 
\end{equation}
\begin{equation}
    \centering
    TNS\_norm  = TNS / clock\_period \
\end{equation}
\begin{equation}
    \centering
     Cong = avg\_cong\_hor + avg\_cong\_ver 
\end{equation}
\begin{equation}
\centering
    cost  = w_{l} \cdot WL\_norm - w_{t} \cdot TNS\_norm + w_{c} \cdot Cong
\end{equation}
\noindent
\textcolor{black}{
where wirelength ($WL\_norm$), TNS ($TNS\_norm$) and congestion ($Cong$) are all collected after clock tree synthesis, $core.width$ and $core.height$ are respectively width and height of the fixed outline (Section \ref{sec:macro_placement}), 
and $avg\_cong\_hor$ and $avg\_cong\_ver$ are respectively the horizontal and vertical congestion reported by Cadence Innovus 21.1.\footnote{
\textcolor{black}{
During the autotuning process, for each configuration ($\phi$, $\zeta$), 
we run {\em Hier-RTLMP} to generate \textcolor{black}{a} macro placement,
and follow this with standard-cell placement
and \textcolor{black}{clock tree synthesis}
using Cadence Innovus 21.1.}}}
\textcolor{black}{
$w_l$, $w_t$ and $w_e$ are corresponding weights to
adjust the values of $WL\_norm$, $TNS\_norm$ and $Cong$ 
so that they contribute equally to the final cost.
Specifically, for Ariane (NG45), $w_l$, $w_t$ and $w_e$ are
respectively set to 1.0, 0.1 and 1.0, 
while for CA53 (GF12), the corresponding values are respectively set to 100, 0.1 and 100.}
\textcolor{black}{
The number of trials allowed to Tune affects QoR: more trials
achieve better QoR at the cost of longer tuning time. 
In our experiments, we set the number of trials to 10 and 50, denoted as $autotune_{10}$
and $autotune_{50}$ respectively.
We use 5 threads to obtain an acceptable tuning walltime 
equal to 2 ($autotune_{10}$) or 10 ($autotune_{50}$) times that of a single 
{\em Hier-RTLMP} run, without any undue CPU needs. 
The best configurations \textcolor{black}{with respect to loss function} found by $autotune_{10}$ and $autotune_{50}$,
noted as $autotune_{10}^*$ and $autotune_{50}^*$ respectively,
are shown in Table \ref{tab:ablation}.
Figure \ref{fig:autotune} shows the design space explored by $autotune_{50}$.
The corresponding \textcolor{black}{best} postRouteOpt layouts are presented in Figure \ref{fig:autotune_layout}.}

\begin{table}[!h]
\caption{
\textcolor{black}{
The effects of ablation study and autotuning for Ariane (NG45) and CA53 (GF12).
We highlight best values of \textcolor{black}{timing} metrics in blue bold font.
Data points for GF12 are normalized.
}
}
\label{tab:ablation}
\resizebox{1\columnwidth}{!} {
\centering
\begin{tabular}{|c|c|c|c|c|c|c|}
\hline
\multicolumn{1}{|l|}{\makecell{ Design\\ (Enablement) }}             
    & \makecell{ Macro\\Placer}
    & \makecell{Std Cell \\ Area ($mm^2$)}   & \makecell{WL \\($m$)}   &  \makecell{Power \\($mW$)}
    & \makecell{ WNS\\($ps$)} &\makecell{TNS \\($ns$) }
    \\ \Xhline{2\arrayrulewidth}
    
\multirow{5}{*}{\makecell{Ariane\\ (NG45) }}                                             
        & {\em default}         
        & 0.247 & 5.27 & 833 & -97 & -55  \\  \cline{2-7}
        & $\phi^* = 3$         
        & 0.247 & 5.27 & 833 & -97 & -55 \\  \cline{2-7}
        & $\zeta^* = 45$         
        & 0.245 & 5.16 & 831 & -99 & -42  \\  \cline{2-7}
        & $autotune_{10}^*$ 
        & 0.247 & 5.28 & 834 
        & \textcolor{blue}{\textbf{-73}} 
        & \textcolor{blue}{\textbf{-29}}  \\  \cline{2-7}
        & $autotune_{50}^*$         
        & 0.247 & 4.89 & 837 & -76 
        & -30  \\ \hline

\multirow{5}{*}{\makecell{CA53\\ (GF12) }}                           
        & {\em default}         
        & 0.43 & 1.08 & 1.03 & -0.26 & -167  \\  \cline{2-7}
        & $\phi^* = 7$         
        & 0.43 & 1.07 & 1.03 & -0.25 & -140  \\  \cline{2-7}
        & $\zeta^* = 15$         
        & 0.43 & 1.08 & 1.03 & -0.26 & -167 \\  \cline{2-7}
        & $autotune_{10}^*$         
        & 0.43 & 1.08 & 1.03 & -0.26 & -167  \\  \cline{2-7}
        & $autotune_{50}^*$         
        & 0.43 & 1.07 & 1.03 & \textcolor{blue}{\textbf{-0.25}} & 
        \textcolor{blue}{\textbf{-140}} \\ \hline
\end{tabular}
}
\end{table}

\subsection{Runtime Analysis}
\label{sec:runtime}

\begin{figure}[!htb]
    \centering
    \includegraphics[page=1, width=0.82\linewidth]{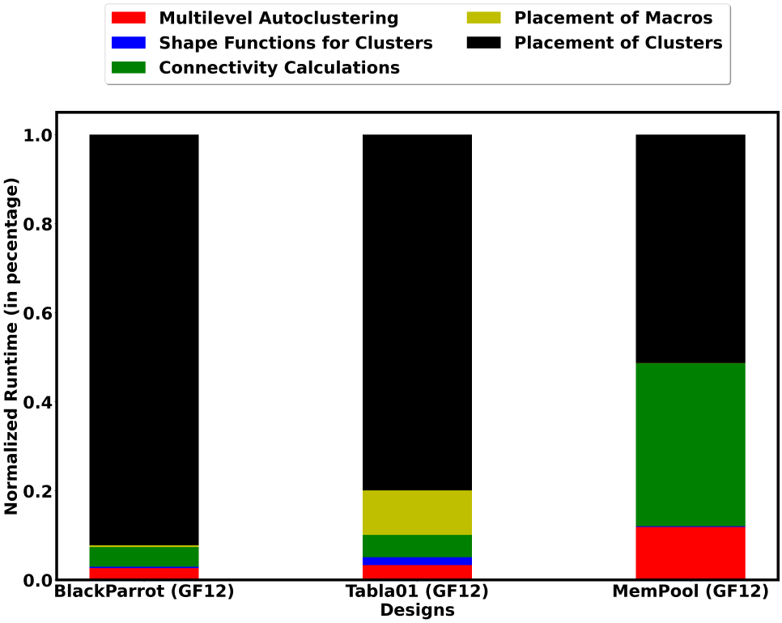}
    \caption{\textcolor{black}{Runtime profiling of {\em Hier-RTLMP} on  BlackParrot (GF12), Tabla01 (GF12) and MemPool (GF12).}}
    \label{fig:runtime}
\end{figure}

\textcolor{black}{
From Table \ref{tab:rtlmp_result}, we can see that {\em Hier-RTLMP} is much faster 
than {\em RTL-MP}.
The profiling result for our macro placer on the Tabla01 (GF12) is
shown in Figure \ref{fig:runtime}.  We can see that
the runtime of sweeping the target dead space $t\_dead\_space$ 
and target utilization $util$ (Section \ref{sec:shape_function})
is not significant.}

\section{CONCLUSION AND FUTURE WORK}
\label{sec:conclusion}

\textcolor{black}{
In this work, we propose {\em Hier-RTLMP}, which extends {\em RTL-MP}  to support multilevel physical hierarchy and hierarchical macro placement.}
Extensions to {\em Hier-RTLMP} \textcolor{black}{that} we are currently 
exploring \textcolor{black}{include}
(i) more intelligent selection of intermediate modules 
of the logical hierarchy as
hierarchical physical clusters based on dataflow and functionality,
\textcolor{black}{
as well as user tagging of logical modules as hierarchical clusters;}
(ii) \textcolor{black}{
fine tuning of the utilization setting for standard-cell clusters
based on the structural complexity, to reduce sweeps and runtime;}
(iii) autotuning \textcolor{black}{of} the cost function based 
on the level of the physical hierarchy and the contents of clusters, 
to better preserve dataflow;
\textcolor{black}{(iv)
enhancing the autoclustering engine to simultaneously
identify the appropriate groupings of macros
and the clusters of standard cells that ``stay together'' through the backend flow;}
and (v) addition of new infrastructure to automatically backtrack up the hierarchy 
to redo a floorplan placement at a higher-level physical cluster, based on 
metrics from a lower-level physical cluster.

\smallskip

\noindent {\textbf{Acknowledgments.}} 
Research at UCSD is partially supported by DARPA IDEA HR0011-18-2-0032
and RTML FA8650-20-2-7009.

\vspace{-1.0cm}

\begin{IEEEbiography}
[{\raisebox{0.3in}{\includegraphics[width=1in, height=1in, clip, keepaspectratio]
{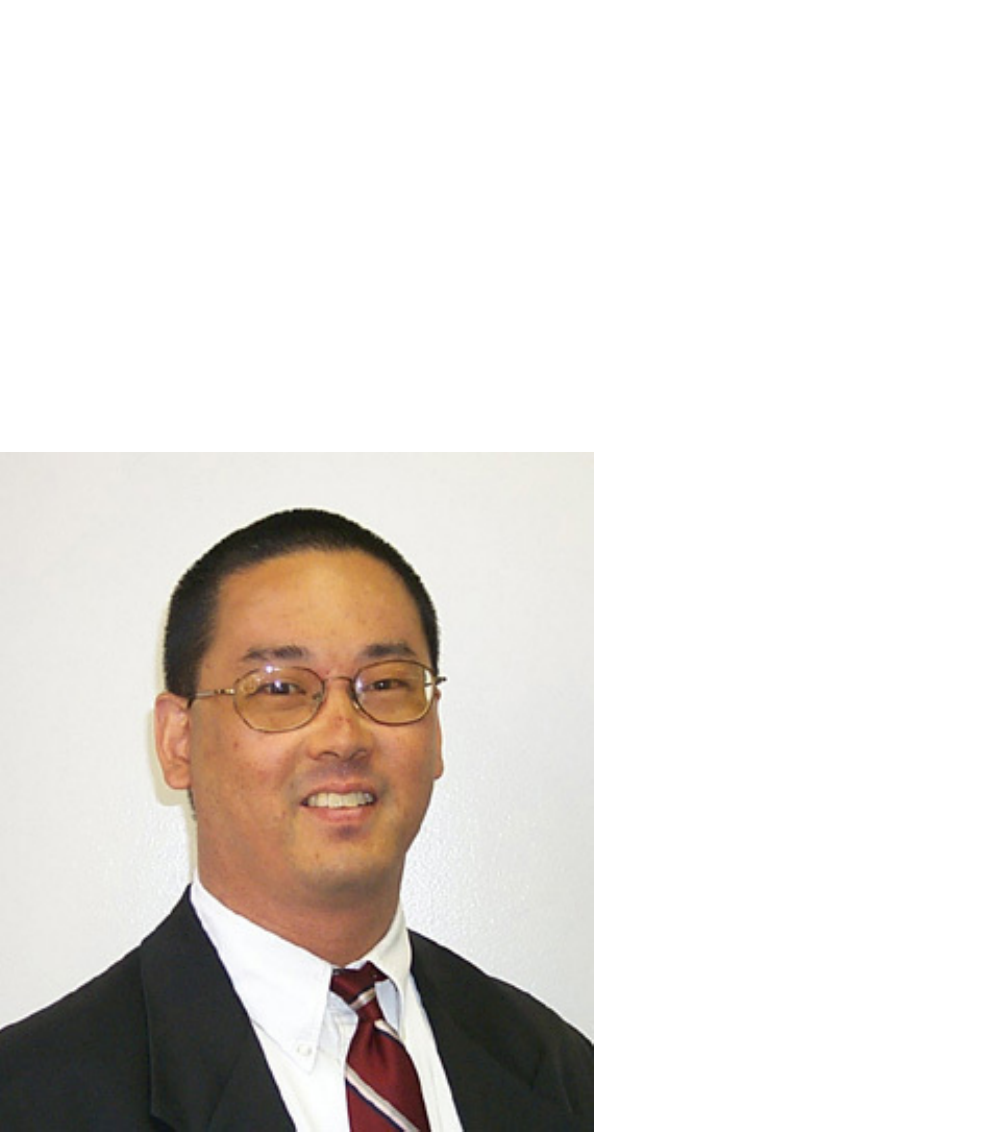}}}]
{Andrew B. Kahng} is a distinguished professor in the CSE and ECE Departments of the University of California at San Diego. His interests include IC physical design, the design-manufacturing interface, large-scale combinatorial optimization, AI/ML for EDA and IC design, and technology roadmapping. He received the Ph.D. degree in Computer Science from the University of California at San Diego.
\end{IEEEbiography}

\vspace{-1.65cm}

\begin{IEEEbiography}
[{\raisebox{0.4in}{\includegraphics[width=0.9in, height=1in, clip,keepaspectratio]
{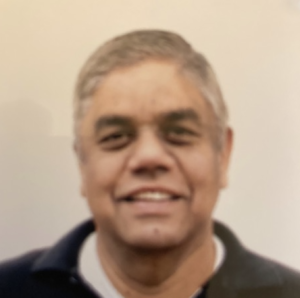}}}]{Ravi Varadarajan} 
is currently pursuing the Ph.D. degree at the University of California 
at San Diego, La Jolla. His research
interests include physical design implementation and methodologies.
\end{IEEEbiography}

\vspace{-2.0cm}

\begin{IEEEbiography}
[{\raisebox{0.4in}{\includegraphics[width=0.9in, height=1in, clip,keepaspectratio]
{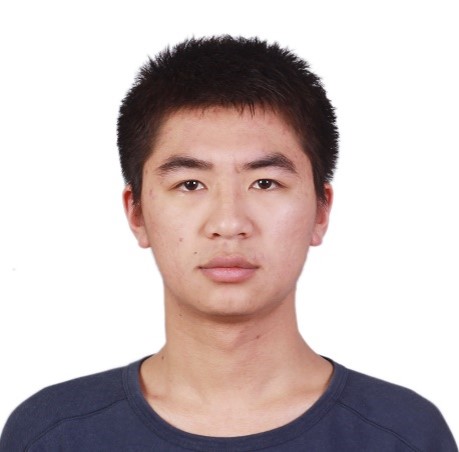}}}]
{Zhiang Wang}
received the M.S. degree in electrical and computer
engineering from the University of California at San Diego, La Jolla, 
in 2022. He is currently pursuing the Ph.D. degree at the University of California 
at San Diego, La Jolla. His current research interests include 
partitioning, placement methodology and optimization.
\end{IEEEbiography}

\end{document}